\documentclass[lettersize,journal]{IEEEtran}
\usepackage{amsmath, amsfonts, algorithmic, array, textcomp, url, verbatim, graphicx, xcolor, booktabs, float, cite, etoolbox, makecell}
\usepackage[caption=false,font=normalsize,labelfont=sf,textfont=sf]{subfig}
\usepackage[noabbrev]{cleveref}
\usepackage[export]{adjustbox}

\def\BibTeX{{\rm B\kern-.05em{\sc i\kern-.025em b}\kern-.08em
    T\kern-.1667em\lower.7ex\hbox{E}\kern-.125emX}}
\usepackage{balance}

\begin{document}

\title{Motional-Current-Sensing Method and Simplified Closed-Loop Control Strategy for Piezoelectric-Resonator-based DC-DC Converters}

\author{Zhiyang Cao,~\IEEEmembership{Graduate Student Member,~IEEE,} Linbo Shao,~\IEEEmembership{Member,~IEEE,} Liyan Zhu,~\IEEEmembership{Member,~IEEE}

\thanks{}}

\markboth{Journal of \LaTeX\ Class Files,~Vol.~18, No.~9, September~2020}
{How to Use the IEEEtran \LaTeX \ Templates}

\maketitle

\begin{abstract}
Piezoelectric resonators (PRs) have been seen as a competitive alternative to magnetic components. In PR-based converters, the motional current (in the LC series branch of the equivalent circuit) is vital for control proposes but cannot be measured directly. The difficulties to detect the zero-crossing points or to measure the amplitude of the motional current has been one of the most dominating obstacles that complicates the control strategies and limits the frequency range of the PR-based converters. This work discusses a ring-dot shaped piezoelectric transformer (PT) based motional current sensing method that provides current information with low-delay, low-loss and intrinsic isolation. It is physically proven that the proposed method is robust with various non-ideal factors of the piezoceramic and circuit implementation. Based on this, an event-driven control strategy is introduced, consisting of only a finite state machine, a PI loop, a low-speed ADC and several comparators. Experiments on a step-down PR-based converter verify that the proposed approach realize ZVS for all transitions within a switching cycle with reduced hardware and software resources, enhances stability and is capable of self-startup.

\end{abstract}

\begin{IEEEkeywords}
piezoelectric resonator, piezoelectric transformer, feedback control, DC-DC converter
\end{IEEEkeywords}

\section{Introduction}

\hbadness=3000

\begin{figure}[h]
    \centering
    \includegraphics[width = 0.7\linewidth]{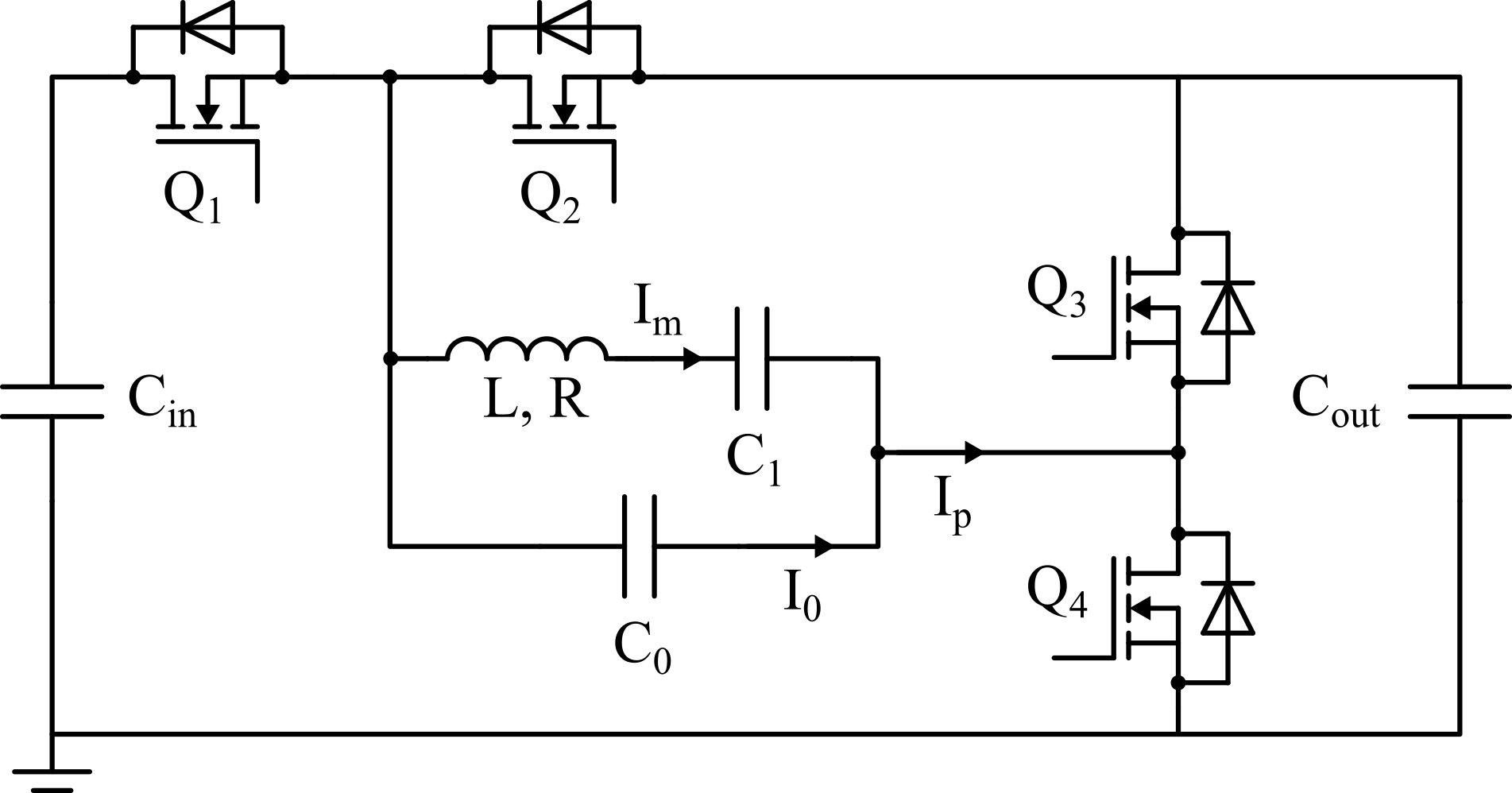}
    \caption{PR-based step down topology.\label{fig:pr-topology}}
\end{figure}

\IEEEPARstart{P}{iezoelectric} components are a compelling alternative to magnetic components in power converters for their high power density, thin profile and better potential to scale down and miniaturize~\cite{bolesOpportunitiesProgressChallenges2023a}. This type of  components utilize the piezoelectric effect and inverse piezoelectric effect to couple electrical energy and mechanical energy. Their application in energy conversion dates to the early 20th century, but back then they often worked as resonant tanks together with conventional magnetic components. In recent years, piezoelectric resonator (PR) based magnetic-less topologies have been proposed~\cite{bolesEnumerationAnalysisDC2021a}. The topologies have been implemented on multiple prototypes~\cite{
bolesEnumerationAnalysisDC2021a,
bolesPiezoelectricResonatorBasedDCDC2023a,
liuSeriesParallelMagneticLess2023a,
stoltFixedFrequencyControlPiezoelectric2021,
stoltCurrentModeControl2025a,
koHybridPiezoelectricResonatorbased2026,
liuClosedloopControlDualside2025a,
liuIntegratedDualSideSeries2024a,
touhamiPhaseShiftVoltageRegulation2024,
touhamiClosedLoopControlSymmetric2026,
pielFeedbackControlPiezoelectricResonatorBased2021a,
yangResonantCurrentEstimation2020a,
forresterBidirectionalInvertingPiezo2023a,
pereiraOperatingFrequencyPrediction2022,
marquesFirst6MHz2025,
bigotNewClosedloopRegulation2024} 
and was used to demonstrate that a single PR can support output power of 1kW with power density of 5.7${\rm kW/cm^3}$~\cite{stoltSpuriousFreePiezoelectricResonator2024}. One of the topologies is shown in Fig.~\ref{fig:pr-topology}, which is capable of step-down DC/DC conversion with $V_{out} < V_{in}/2$.

However, a notable portion of the reported prototypes operates without closed-loop control
~\cite{bolesPiezoelectricResonatorBasedDCDC2023a,
bolesEnumerationAnalysisDC2021a,
koHybridPiezoelectricResonatorbased2026,
liuSeriesParallelMagneticLess2023a,
stoltFixedFrequencyControlPiezoelectric2021,
stoltSpuriousFreePiezoelectricResonator2024}. 
This is largely because the PR-based non-isolated topologies require far more complicated switching sequences and closed-loop control than magnetic-based converters. For adjustable conversion ratio and minimal loss, the control system often involve a switching sequence of six or more stages and complicated constraints~\cite{bolesEnumerationAnalysisDC2021a}.

Researchers have proposed various strategies and implementations, but a majority of them either relies on costly hardwares like high-speed ADCs or requires sophisticated software resources of high-performance controllers. One of the major difficulties from the perspective of control is that the {\it ``motional current''} ($I_m$ in Fig.~\ref{fig:pr-topology}) cannot be directly measured. This quantity helps determine the transition between different switching stages~\cite{bolesEnumerationAnalysisDC2021a} and can be utilized for current-mode control~\cite{stoltCurrentModeControl2025a}. The zero-crossing points of the motional current are especially of interest because the stages can be constraint to end with them to eliminate charge circulation and improve efficiency~\cite{bolesEnumerationAnalysisDC2021a}. However, $I_m$ is only an imaginary representation of the displacement of charges within the material induced by the inverse piezoelectric effect. The externally observed current is a composition of $I_m$ and the current flowing through the static capacitance, $I_0$, which cannot be simply separated. 

As a result, the existing control strategies rely on various estimation method to detect the zero-crossing points of $I_m$ or to bypass the problem. They can be categorized into:

\begin{itemize}
     \item [(a)] Detecting the zero-crossing points of the motional current by resistors or diodes in series with the PR~\cite{yangResonantCurrentEstimation2020a, forresterBidirectionalInvertingPiezo2023a,touhamiImplementationControlStrategy2020} with comparators, while introducing additional loss. PLLs may be required for stability~\cite{yangResonantCurrentEstimation2020a, forresterBidirectionalInvertingPiezo2023a}.
    
    \item [(b)] Deducing the zero-crossing points of the motional current by the ``geometrical'' features of the voltage waveform of the PR, which must be extracted within very tight sampling windows by high-speed ADCs~\cite{pielFeedbackControlPiezoelectricResonatorBased2021a,liuClosedloopControlDualside2025a}. PLLs are also needed in some cases~\cite{liuClosedloopControlDualside2025a}.
    
    \item [(c)] Deducing the zero-current points by monitoring the rising time of the voltage across PR within certain switching stage.~\cite{touhamiPhaseShiftVoltageRegulation2024,touhamiClosedLoopControlSymmetric2026}. This method is adopted together with a balanced eight-stages switching sequence, with which the seven distinct time spans to be controlled in six-stage sequences can be reduced to four. However, the existing approaches still require high-speed ADC sampling~\cite{touhamiPhaseShiftVoltageRegulation2024,touhamiClosedLoopControlSymmetric2026}. Moreover, the inclusion of an additional stage within a single switching period may further shorten the duration of one or more individual stages, which poses more challenges for high frequency operation.
    
    \item [(d)] Using real-time computation to calculate the optimal operating frequency~\cite{marquesFirst6MHz2025, bigotNewClosedloopRegulation2024, pereiraOperatingFrequencyPrediction2022}. Recent progresses have reduced the computational load and demonstrated the feasibility to build closed-loop systems for 6 MHz piezoelectric DC/DC converters with this method and the aforementioned eight-stag sequences~\cite{marquesFirst6MHz2025}. Yet the computation still relies on mainstream MCUs or DSPs. It should also be noted that the controller may not be able to update the control parameters cycle-by-cycle, but at a lower frequency instead. The latency may result in right-half-plane zeros (RHPZs) and have negative effects on robustness.
    
    \item [(e)] Measuring the amplitude of the motional current during specific periods within a switching cycle when the current in the static branch is negligible, for current mode control~\cite{stoltCurrentModeControl2025a}. The solution is straightforward but is only capable of tracking the motional current during a portion of a cycle. Besides, the conventional current-sensing introduces resistors in series with the PR and causes loss.
\end{itemize}

It is clear that a more physical measurement of the motional current can significantly ease the burden of the control system while increasing the robustness. In this work, we analyze the feasibility of using the ring-dot shaped PTs under radial extensional (RE) mode to measure the motional current and propose a novel simplified control strategy with current information. Section II derives the electromechanical model of ring-dot PTs and proves the feasibility of motional-current-sensing. Section III introduces the control strategy. The electromechanical model and the control strategy is verified on an experimental platform in section IV.

\section{PT Based Motional-Current-Sensing}

\hbadness=3000

Since the motional current originates in the mechanical vibration of the piezoceramic, an intuitive approach is to measure the vibration itself. Although direct measurement would be impractical, it is still feasible to transmit the acoustic wave from one piece of piezoceramic to another and measure the induced voltage on the additional one. This essentially forms a piezoelectric transformer (PT), of which the primary side is connected to the power loop as an ordinary PR and the secondary side is connected to certain downstream circuits that restore the primary side motional current from the secondary side induced voltage. The approach is named as \textit{``Motional-Current-Sensing''} in this paper, and the transfer function between the primary side motional current and the secondary side voltage is referred to as the \textit{``current-sensing transfer function''}.

It should be noted that despite the transformer-like configuration, the component proposed in this work does not mean to route \textit{energy} between two isolated ports, but \textit{signals}, or \textit{information} instead. However, PTs can also be used to transmit power from one port to the other in power electronics converters\cite{navalHighEfficiencyIsolatedPiezoelectric2026,ngPiezoelectricTransformerComponent2023a,wangPlanarPiezoelectricTransformerBased2022a}, surfacing acoustic wave filters~\cite{morganSurfaceAcousticWave2010} oscillators~\cite{liLowPhaseNoise2020} and sensors~\cite{xiRoomTemperatureMidInfraredDetection2025}.

\begin{figure}[h]
    \centering
    \includegraphics[width = 0.8\linewidth]{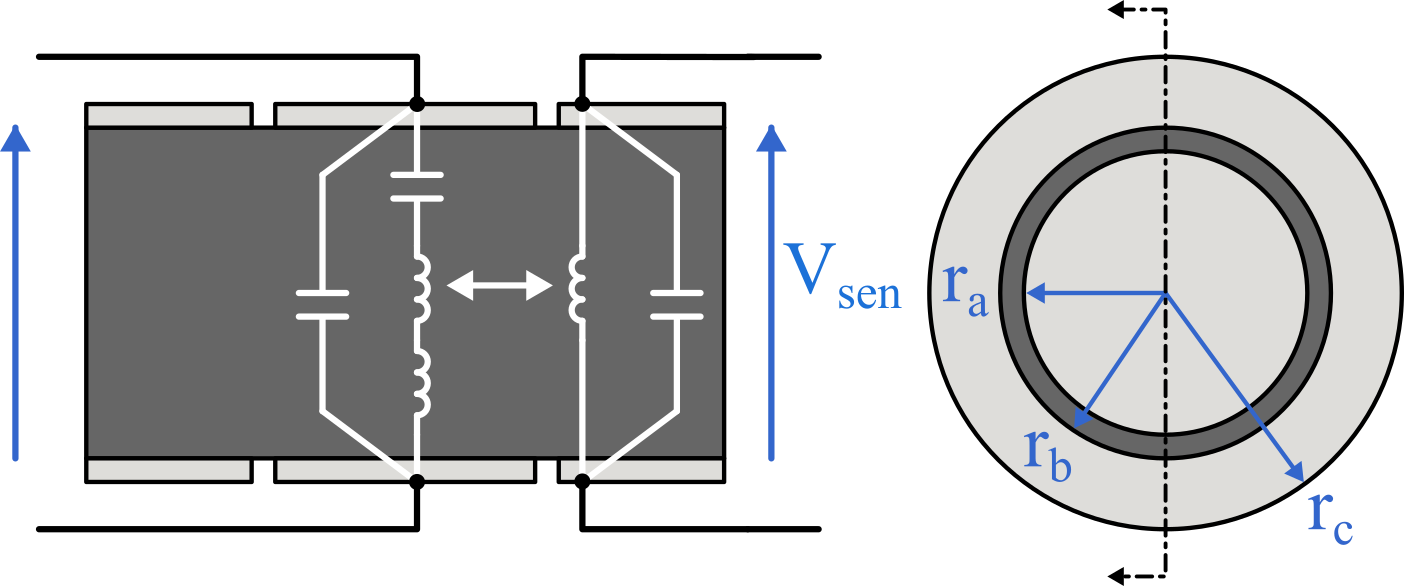}
    \caption{A ring-dot PT and its simplified equivalent circuit. The illustration is not to real scale.\label{fig:piezo-ring-dot}}
\end{figure}

The structure of the current-sensing PT should be determined by the type of the original PR. Most of the reported piezoelectric converters utilized disk-shaped PRs. For a piezoceramic plate with $d >> h$, the transverse effect (T-effect) is dominant at low frequency of several hundred kHz, with which the strain generated via inverse piezoelectric phenomenon is perpendicular to the polarization direction, leading to the RE mode~\cite{dingAccurateCoupledVibration2021,
guoMeasurementPredictionFrequency1992a,huangTheoreticalAnalysisExperimental2004}. To transmit the vibration under RE mode, a radially nested structure is the most appropriate, forming a namely ring-dot PT in Fig.~\ref{fig:piezo-ring-dot}. The ``dot'' section in the center and the ``ring'' section at the edge are covered with conductive electrodes on both sides, while the middle section that separates them is not. In this way, the inner section can work as the primary side of the PT and the outer section can be the isolated secondary side, and the mechanical vibration of both ports are coupled through the middle section.

\begin{figure}[h]
    \centering
    \includegraphics[width = 0.8\linewidth]{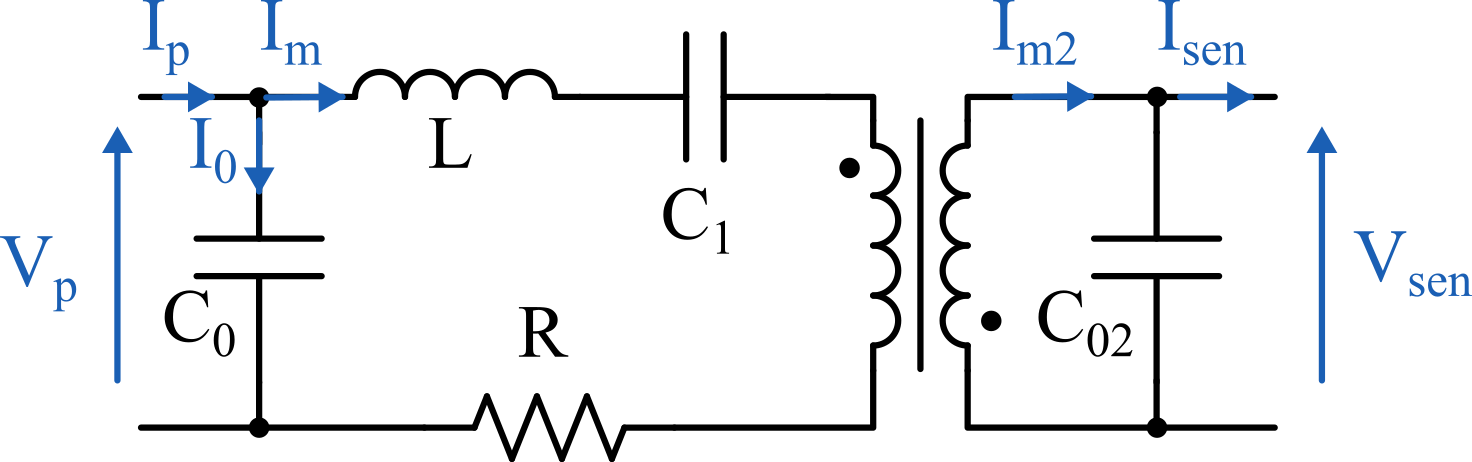}
    \caption{Reduced Mason model of piezoelectric transformers around one of the resonant frequencies.\label{fig:mason-model}}
\end{figure}

In this section, the theoretical feasibility of utilizing the PT for motional current sensing is analyzed. Two quantities are of interest in particular: First, the phase between the sensing voltage and the motional current must be a known value and be stable over the range of operating frequency. This is a strict requirement for zero-crossing point detection; Second, the amplitude of the sensing voltage should also follow the motional current so that it can be used for current-mode control. This, however, is generally allowed to fluctuate since the errors can be compensated by the control system.

The physical quantities used in this paper are listed in Table~\ref{tab:physical-quantity}. 

\begin{table}[h]
    \centering
    \caption{Physical Quantities and Explanations}
    \begin{tabular}{ll}
        \toprule
        Quantity & Explanation \\
        \midrule
        $V_{p}, I_{p}$ & Primary side voltage \& current\\
        $V_{sen}, I_{sen}$ & Secondary side voltage \& current \\
        $I_{m}, I_{m2}$ & Motional current \\
        $v_{[a,b,c]}r_{[a,b,c]}$ & Area expansion rate \\
        $C_{0}, C_{02}$ & Static capacitances \\
        $2C_{lk}$ & Leakage capacitance \\
        $Z_{[i,m,o][1,2,3]}$ & Mechanical impedances \\
        $A$ & Conversion ratio between motional current and vibration \\
        $C_x, L, R$ & Equivalent circuit components \\

        $r_{[a,b,c]}$ & Radiuses of different sections \\
        $h$ & Component thickness \\

        $s^{x}$ & Compliance tensor \\
        \hspace{0.2cm} $s^{E}$ & \hspace{0.2cm} under constant electric field \\
        \hspace{0.2cm} $s^{D}$ & \hspace{0.2cm} under constant electric displacement field \\

        ${\sigma^x}$ & Poisson's Ratio \\
        \hspace{0.2cm} ${\sigma^E}$ & \hspace{0.2cm} under constant electric field \\
        \hspace{0.2cm} ${\sigma^D}$ & \hspace{0.2cm} under constant electric displacement field \\

        $d$ & Piezoelectric coefficient tensor \\
        $\rho$ & Mass density \\
        \bottomrule
    \end{tabular}
    \label{tab:physical-quantity}
\end{table}

\subsection{Steady-State Current-Sensing Transfer Function under Ideal Condition}

The reduced Mason model~\cite{masonPhysicalAcousticsProperties1958} of the piezoelectric transformers in Fig.~\ref{fig:mason-model} can be utilized to analyze the steady-state current-sensing transfer function under ideal condition, that is, without any loss or parasitic phenomenon. $L$ and $C_1$ form the motional branch. $C_0$ and $C_{02}$ are static capacitances between the electrodes of the dot and ring section, respectively. $R$ is the equivalent resistance representing material loss that is ideally zero. The coupling between primary and secondary side motional current $I_m$ and $I_{m2}$ in Fig.~\ref{fig:piezo-ring-dot} can be modelled as an ideal transformer~\cite{ikedaFundamentalsPiezoelectricity1996,navalHighEfficiencyIsolatedPiezoelectric2026}. The resonant frequency of the motional branch corresponds to the mechanical vibration mode. While only the first RE mode is studied in this work, the equivalent circuit in Fig.~\ref{fig:mason-model} can be modified to represent multiple modes by introducing additional motional branches~\cite{navalOvertonePiezoelectricTransformers2025}.

The equivalent circuit intuitively leads to two features of the current-sensing transfer function: The secondary side motional current $I_{m2}$ is always the reverse of $I_m$ of the primary side, thus the phase of the secondary side voltage $V_{sen}$ is 90 degrees ahead of $I_m$ due to the charging and discharging processes of the capacitance $C_{02}$. Besides, the amplitude of $V_{sen}$ is proportional to that of $I_m$ because the system is linear. These observations are concluded in~\eqref{eq:conlusion-under-ideal-condition}.

\begin{subequations}
    \begin{align}
        \phi \left( {\frac{{{V_{sen}}}}{{{I_m}}}} \right) &\equiv \frac{\pi }{2} 
        \\
        \left| {{V_{sen}}} \right| &\propto \left| {{I_m}} \right|
    \end{align}
    \label{eq:conlusion-under-ideal-condition}
\end{subequations}

\begin{figure}[h]
    \centering
    \includegraphics[width = \linewidth]{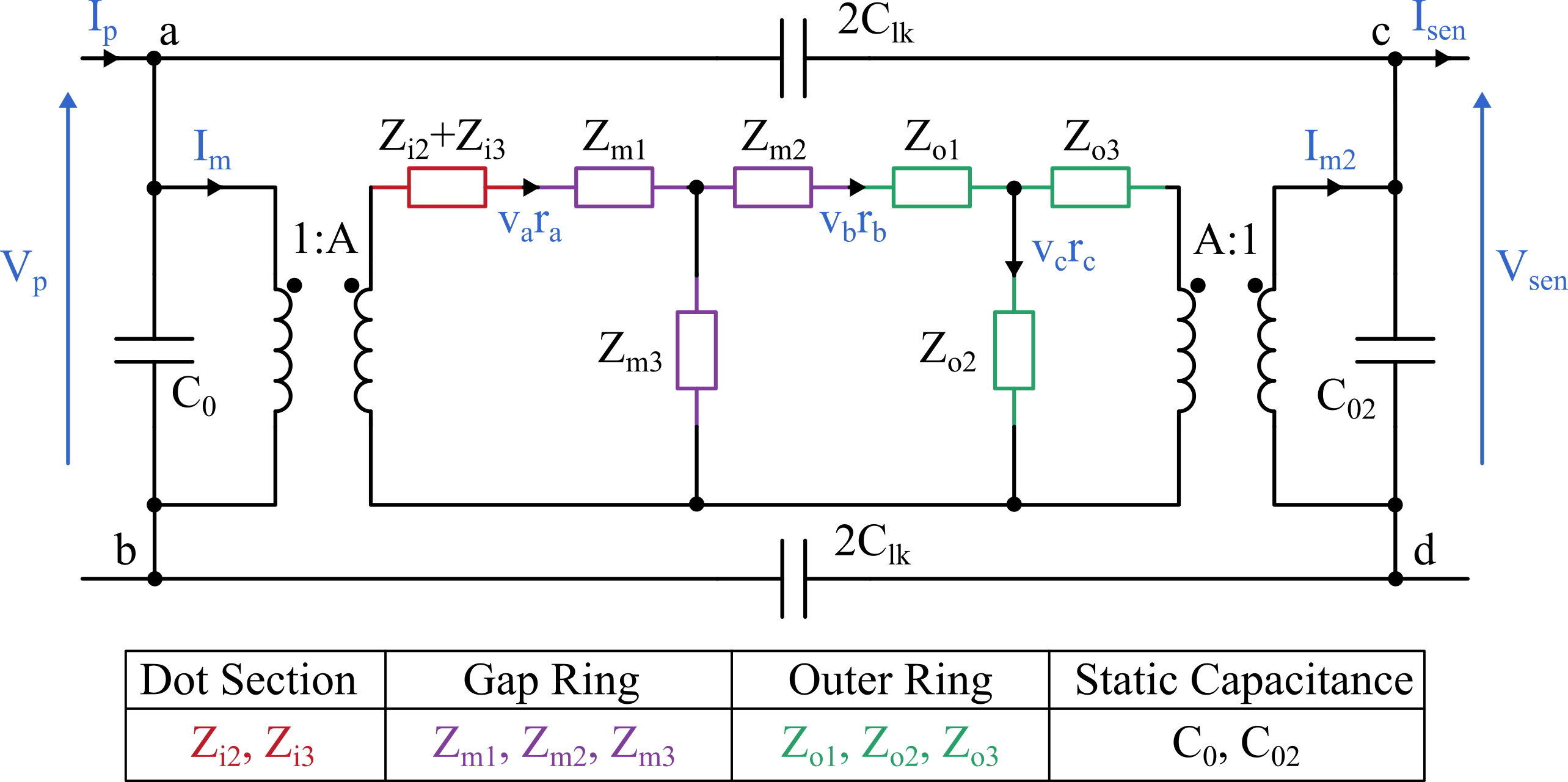}
    \caption{The equivalent electromechanical circuit of ring-dot PTs.}
    \label{fig:ring-dot-equivalent-circuit}
\end{figure}

In simulations and experiments, however, it is observed that although the phase of the current-sensing transfer function is indeed around $90^\circ$ near the resonant frequency, it is not a fixed value but varies with frequency and certain non-ideal factors, which the reduced Mason model cannot explain. Since the phase of the transfer function is crucial in this work, a more fundamental electromechanical model of the ring-dot PT as shown in Fig.~\ref{fig:ring-dot-equivalent-circuit} must be used. In the model, $Z_{[i,m,o][1,2,3]}$ denotes the mechanical impedances of the piezoelectric plate, $A$ is conversion ratio between the motional current and vibration velocity, $C_{lk}$ is the leakage capacitance between the two ports.

Further analysis in this section are based on the complete current-sensing transfer function derived from the electromechanical model in Appendix~\ref{appendix:derivation}.  Fig.~\ref{fig:model-simulation} shows the calculated transfer function and the admittance seen from the primary side near the first resonant frequency. With the non-idealities taken into consideration, the phase of the current-sensing transfer function oscillates between $\pm 90^\circ$ over a wide frequency range, but shows excellent stability between the first resonant frequency $f_{r1}$ and the first anti-resonant frequency $f_{ar1}$. However, due to the non-ideal factors, $V_{sen}$ is $90.22^\circ$ ahead of $I_{m}$ instead of precisely $90^\circ$ at $f_{r1}$. It will be discussed next where this error originates and how to limit its impact.

\begin{figure}[h]
    \centering
    \includegraphics[width = 0.85\linewidth]{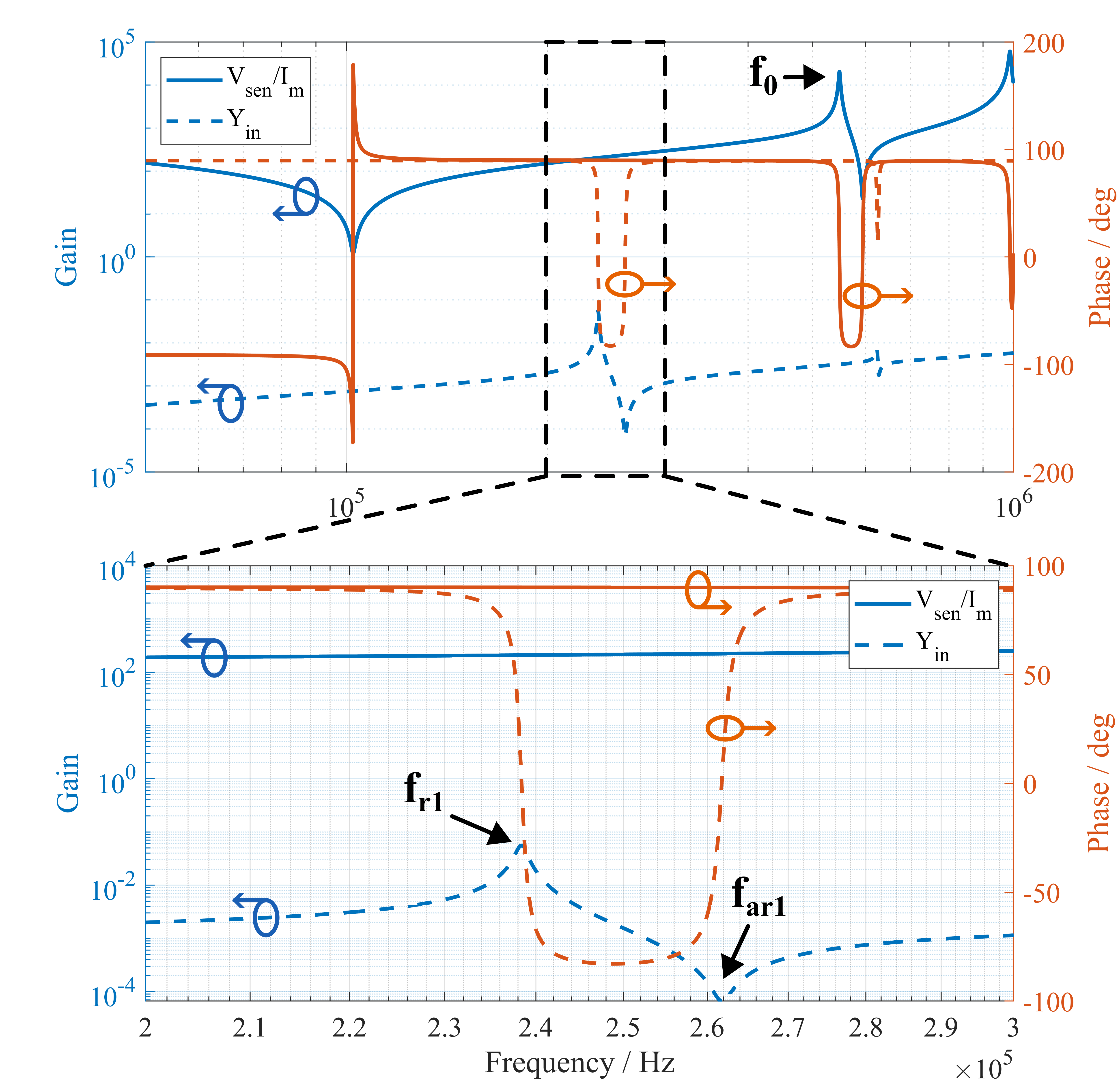}
    \caption{Calculated input admittance of the primary side $Y_{in}$ (solid lines) and the current-sensing transfer function (dotted lines). Calculation uses properties of PZT-8 material and $r_a = 4{\rm mm}$, $r_b = 4.5{\rm mm}$, $r_c = 5{\rm mm}$, $h = 0.5{\rm mm}$, $\delta_m = 0.5\%$, $\delta_e = 0.5\%$.}
    \label{fig:model-simulation}
\end{figure}

\subsection{Non-ideal Factors}

\subsubsection{Material Loss}

Two types of losses in piezoelectric materials are considered here: mechanical loss that dissipates the propagation of the acoustic wave; dielectric loss that dissipates the electrical energy during the swapping of polarization direction. They introduce imaginary component to the stiffness constant $c^E$ as well as the dielectric constant $\varepsilon^T$~\cite{ikedaFundamentalsPiezoelectricity1996}, and consequently introduce real component to the mechanical impedances. As a result, the phase of the current-sensing transfer function in~\eqref{eq:vout-irin} will no longer be a binary function of $\pm \pi/2$, but vary gradually between them.

In this work, the loss factors are simplified as isotropic and homogeneous. Their impact is analyzed by substituting~\eqref{eq:ring-loss} into~\eqref{eq:mechanical-impedance}:

\begin{subequations}
    \label{eq:ring-loss}
    \begin{align}
        {c^{x*}} &= {c^{x}}\left( {1 + j{\delta _c}} \right)\quad {s^{x*}} = {\left( {{c^{x*}}} \right)^{ - 1}}
        \label{eq:ring-loss-ce}
        \\
        {\varepsilon ^{T*}} &= {\varepsilon ^T}\left( {1 - j{\delta _\varepsilon }} \right)
        \label{eq:ring-loss-epsilon}
    \end{align}
\end{subequations}

\begin{figure}[h]
    \centering
    \includegraphics[trim=11mm 12mm 15mm 5mm, clip, width = 0.9\linewidth]{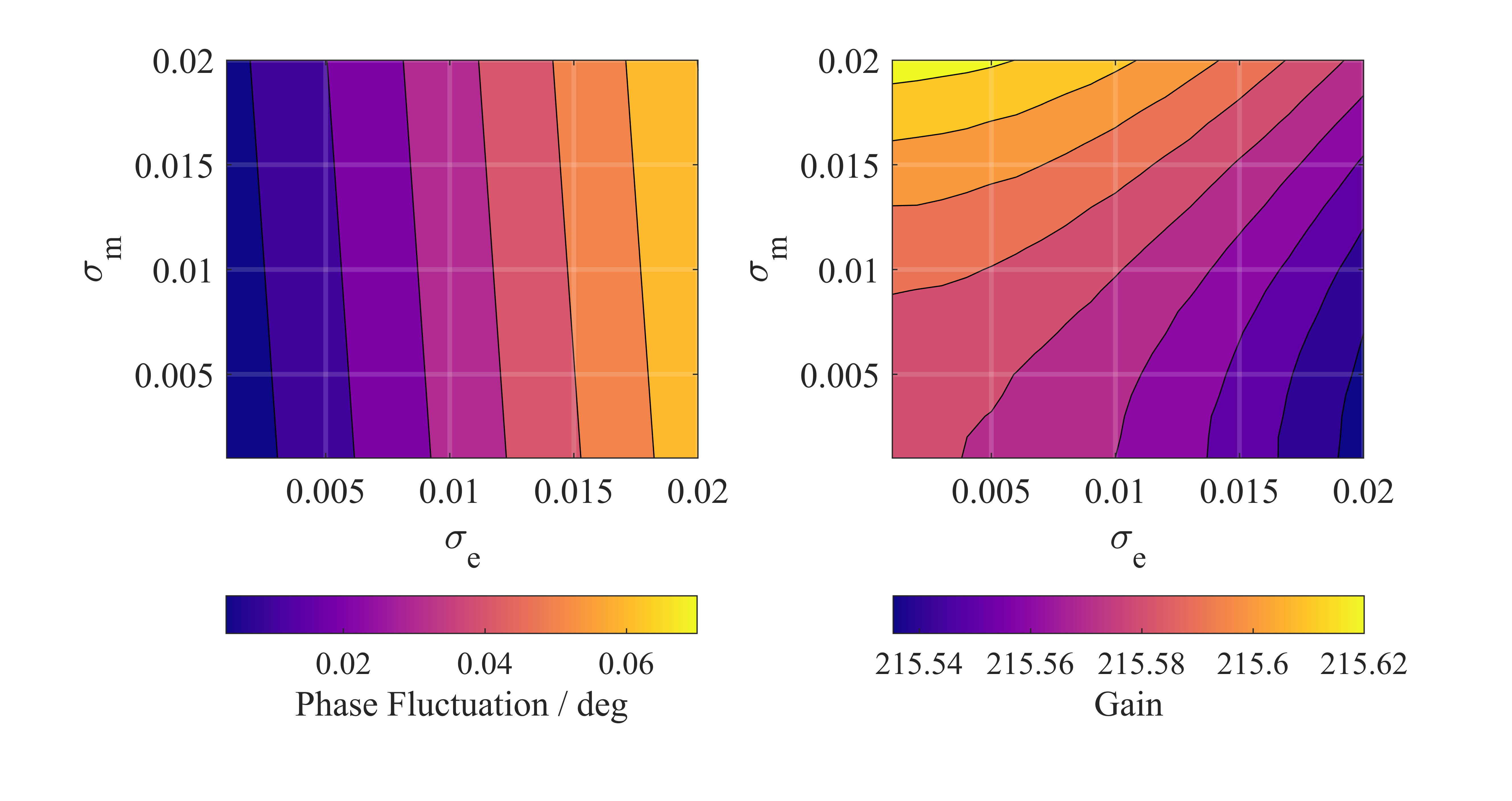}
    \caption{Calculated current-sensing transfer function at resonant frequency versus loss factors ${\sigma^x}_m$ and ${\sigma^x}_e$. $r_a = 4{\rm mm}$, $r_b = 4.5{\rm mm}$, $r_c = 5{\rm mm}$, $h = 0.2{\rm mm}$.\label{fig:loss-sweep}}
\end{figure}

In Fig.~\ref{fig:loss-sweep}, the impact of loss factors are numerically calculated and evaluated by the gain of the current-sensing transfer function at the first resonant frequency and the \textit{fluctuation range of the phase} between the first resonant frequency and the first anti-resonant frequency, which is defined as~\eqref{eq:phase-error}.
\begin{equation}
    {\delta _\phi } = \max \phi  - \min \phi \quad ,{\omega _{r1}} < \omega  < {\omega _{ar1}}
    \label{eq:phase-error}
\end{equation}

It is observed that the dielectric loss affects the phase error more significantly than the mechanical damping, while the latter is more correlated to the amplitude of the current-to-voltage gain. 

\subsubsection{Leakage Capacitance}
The electrodes on the same side of the PT form a coaxial-coplanar capacitor creating a leakage path between the input and output terminals. Each of the capacitance is denoted as $2C_{lk}$ in Fig.~\ref{fig:ring-dot-equivalent-circuit} so that the two on both sides can be merged into $C_{lk}$ for simplicity of discussion. The leakage capacitance is positively correlated to the thickness of the component and negatively correlated to the width of the middle section ($r_b - r_a$), but the exact value is difficult to obtain because of the sophisticated distribution of electric field near the edge of adjacent sections.

The impact of $C_{lk}$ on the current-sensing transfer function can be first analytically studied without material loss using the method in Appendix~\ref{appendix:derivation}. It can be derived that $C_{lk}$ in the complete current-sensing transfer function will flip the phase to $-90^\circ$ and flip back to $+90^\circ$ as the frequency rises. The second flipping point approaches the first resonant frequency $f_{r1}$ from the low-frequency side as $C_{lk}$ increases, but can never reach $f_{r1}$. Thus it is guaranteed that the phase of the current-sensing transfer function is still $+90^\circ$ at the resonant frequency. However, with material loss, the phase cannot change abruptly, but varies continuously. As a result, the phase of the current-sensing transfer function at $f_{r1}$ will be lower with larger $C_{lk}$, and its fluctuation range between $f_{r1}$ and $f_{ar1}$ will be more significant.

\begin{figure}[h]
    \centering
    \includegraphics[width = 0.85\linewidth]{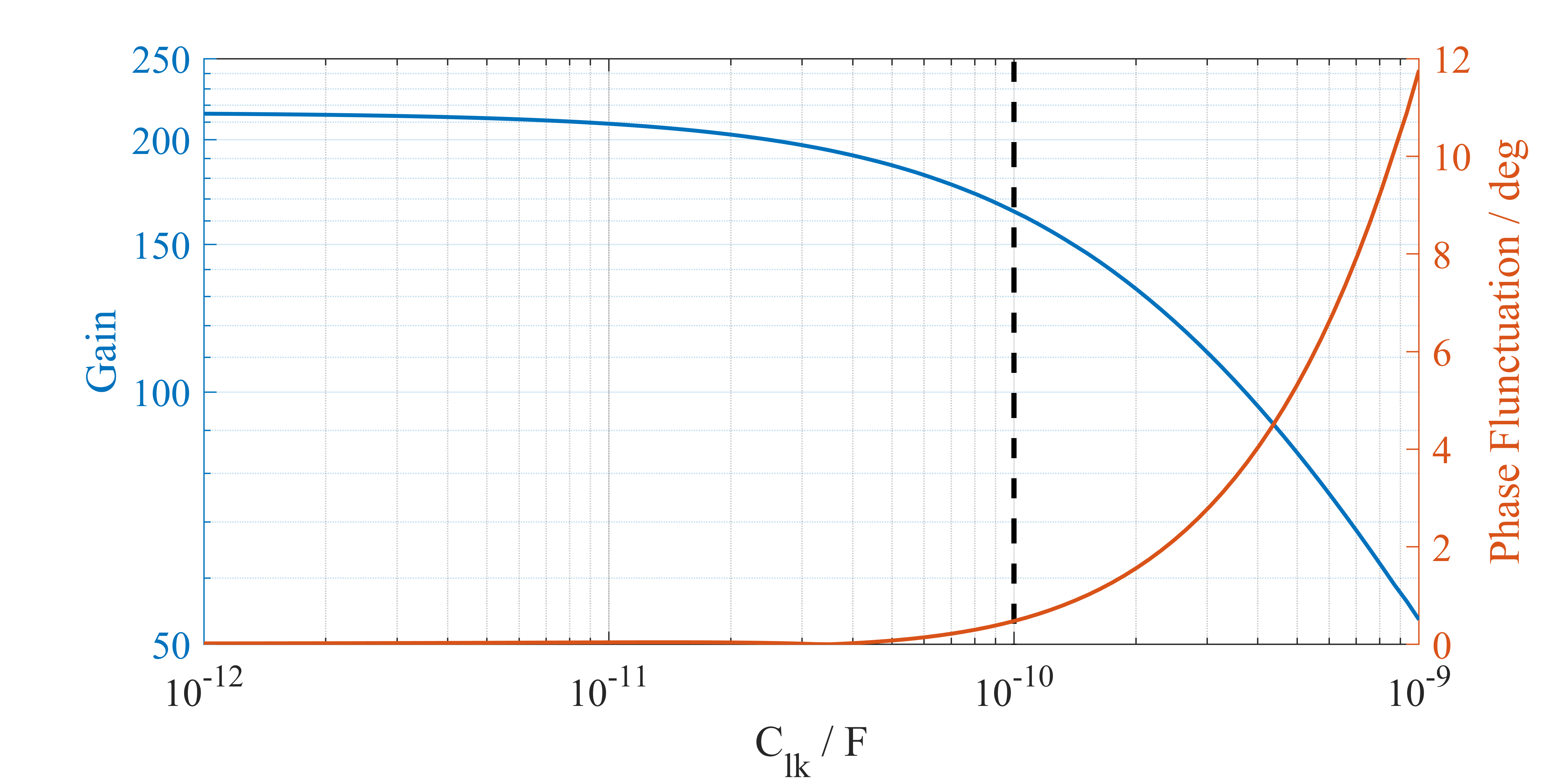}
    \caption{Current-sensing transfer function at resonant frequency versus leakage capacitance $C_{lk}$. $r_a = 4{\rm mm}$, $r_b = 4.5{\rm mm}$, $r_c = 5{\rm mm}$, $h = 0.5{\rm mm}$, $\delta_m = 0.5\%$, $\delta_e = 0.5\%$.\label{fig:clk-sweep}}
\end{figure}

Fig.~\ref{fig:clk-sweep} shows the gain of the current-sensing transfer function at the first resonant frequency and the maximum phase error between the first resonant frequency and the first anti-resonant frequency. For the shown case, the impact of $C_{lk}$ is trivial as long as $C_{lk} < 100 {\rm pF}$, while the simulated value using finite element method (FEM) is $19.4 {\rm pF}$. It is presumable that the leakage capacitance is not a major concern, especially with a reasonable isolation distance between the electrodes ($r_c - r_b$).

\subsubsection{Secondary Side Load}
\begin{figure}[h]
    \centering
    \includegraphics[trim=11mm 12mm 15mm 5mm, clip, width = 0.9\linewidth]{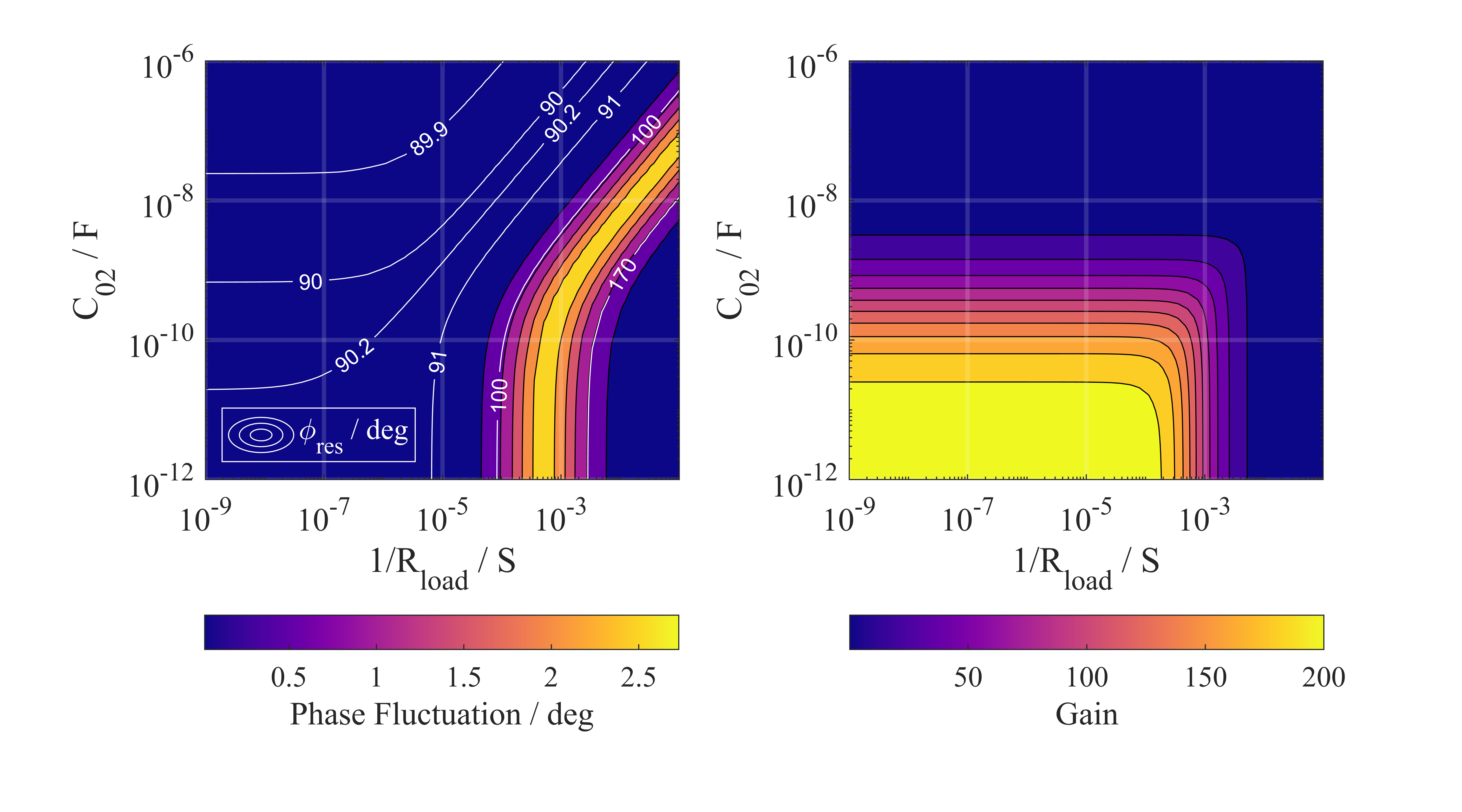}
    \caption{Calculated current-sensing transfer function at resonant frequency versus secondary side load. The absolute phase of the transfer function at resonant frequency is plotted with white curves in the left figure. $r_a = 4{\rm mm}$, $r_b = 4.5{\rm mm}$, $r_c = 5{\rm mm}$, $h = 0.5{\rm mm}$, $\delta_m = 0.5\%$, $\delta_e = 0.5\%$.}
    \label{fig:load-sweep}
\end{figure}

As revealed in~\eqref{eq:vout-irin}, the real component of the output current $I_{sen}$ will result in non-90$^\circ$ phase-shift, whereas the imaginary component does not have such impact unless the loss factors introduce real part to the mechanical impedances. However, numerical solution reveals that the phase fluctuation between the resonant frequency and anti-resonant frequency is barely affected, which is shown in Fig.~\ref{fig:load-sweep}. Here, $C_{02}$ denotes the combination of the intrinsic capacitance of the ring section and all external capacitance in parallel with it.

Fig.~\ref{fig:load-sweep} shows the possibility to adjust the current-to-voltage gain by controlling the secondary side load. Given that the current-to-voltage gain is affected by both real and imaginary load while the phase fluctuation is more resistant to the latter, it is more desirable to adjust with $C_{02}$ and keep the input resistance of the following circuit high. This will be further discussed in tne next subsections.

\subsubsection{Voltage and Heat}
It is of interest whether the current-sensing transfer function derived in Appendix~\ref{appendix:derivation} is still valid with higher voltage amplitude or higher voltage bias. Experiments~\cite{danielNonlinearLossesMaterial2024} have shown that despite observable non-linearity, the electrical impedance of the component remains stable until the voltage approaches the maximum rating of the material, indicating that the loss factors are not significantly altered. It should also be noted that although the loss of PZT degrades as the voltage rises, that of LN is actually better. Therefore, it is presumable that the current-sensing remains accurate over a reasonable range of voltage according to Fig.~\ref{fig:loss-sweep}.

The concern of heat also arises when the power density increases, as the piezoelectric coupling factor and other mechanical properties varies with temperature~\cite{gubinyiElectricalPropertiesPZT2008}. However, the correlation between the electrical properties and temperature varies widely among different materials even within the same type~\cite{danielNonlinearLossesMaterial2024,gubinyiElectricalPropertiesPZT2008}, and is beyond the scope of this work. Nevertheless, the reported data~\cite{gubinyiElectricalPropertiesPZT2008} reveal that the variation of material properties is limited until the temperature approaches the curie temperature, which is $\rm{160^\circ C - 350^\circ C}$ for PZT materials and $\rm{>1,500^\circ C}$ for LN~\cite{zhangPiezoelectricMaterialsHigh2011}. In general, PZT materials is more sensitive to temperature as compared with LN~\cite{danielNonlinearLossesMaterial2024}, but the optimal temperature can either be higher or lower than room temperature.

\subsection{Downstream Circuit}

\begin{figure}[h]
    \centering
    \includegraphics[width = 0.8\linewidth]{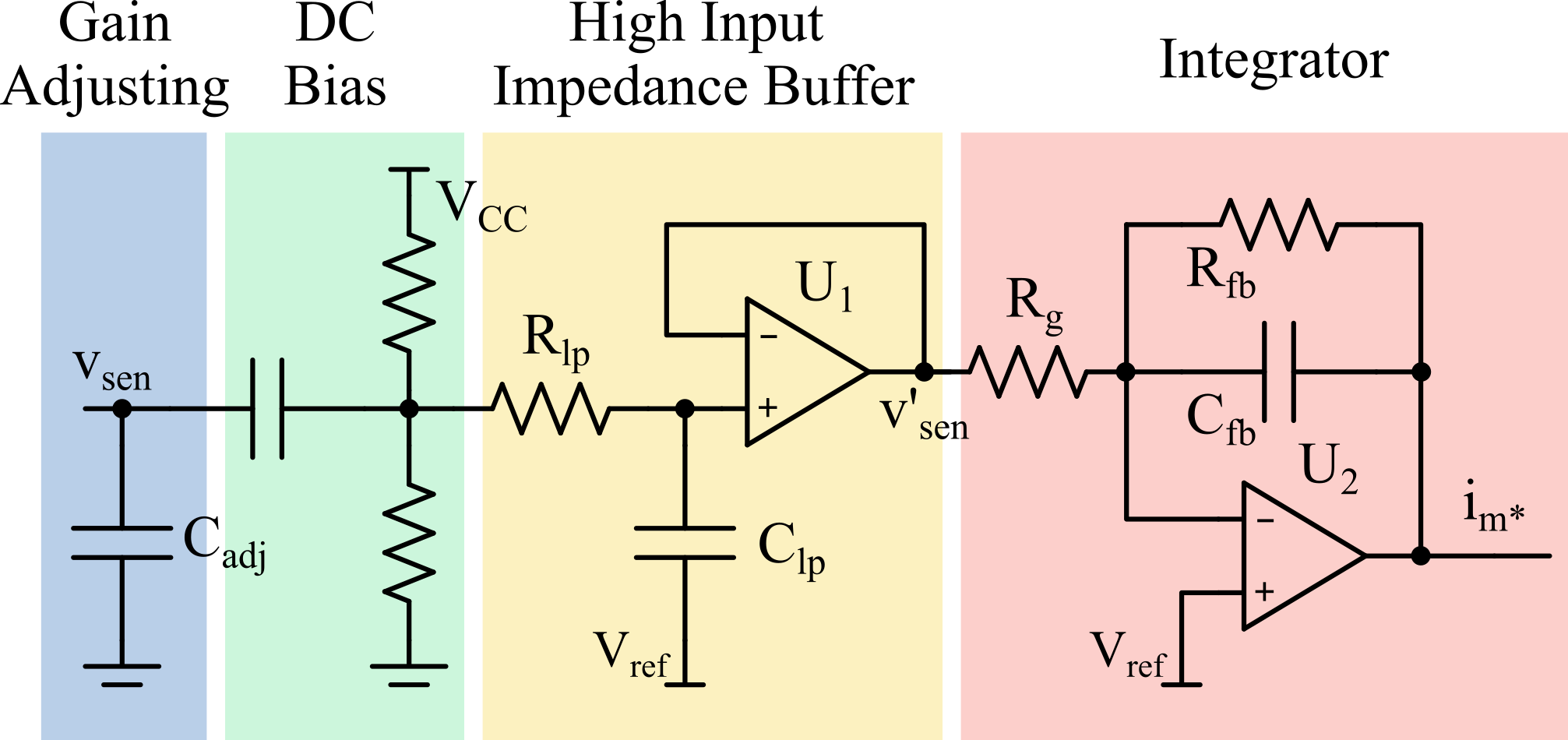}
    \caption{Motional current restoring circuit.\label{fig:phase-shifter}}
\end{figure}

It has been demonstrated that the sensing voltage $V_{sen}$ is $90^\circ$ ahead of the primary side motional current $I_m$ in a current-sensing PT at steady state with reasonable non-ideal factors. A phase-shifting circuit is then required to restore $I_m$ from the $V_{sen}$. As shown in Fig.~\ref{fig:phase-shifter}, the circuit used in this work consists of a high-input impedance buffer and an integrator. $R_{lp}$ and $C_{lp}$ form a low-pass filter, $R_{g}$, $R_{fb}$ and $C_{fb}$ form an inverse integrator with $U_2$. Together, the circuit introduce a $+90^\circ$ phase shift on $V_{sen}$, so the restored current $I_{m*}$ is $180^\circ$ from $I_m$ at steady state.

The transfer function of the phase-shifting circuit writes:
\begin{equation}
    \frac{{{I_{m*}}}}{{{V_{sen}}}} = \frac{1}{{j\omega {R_{lp}}{C_{lp}} + 1}} \cdot \frac{{ - {R_{fb}}/{R_{g}}}}{{j\omega {R_{fb}}{C_{fb}} + 1}}
    \label{eq:phase-shifting}
\end{equation}

The integrator provides $90^\circ \sim 180^\circ$ phase shift while the low-pass filter enables $0^\circ \sim 90^\circ$. By adjusting the components, the phase-shifting circuit is able to handle any $\phi(G)$ greater or less than $90^\circ$. Its design procedure will be discussed later.

Another possible solution to restore the primary side motional current is to short-circuit the secondary side with a shunt resistor and measure the secondary side motional current $I_{m2}$ with it. The reduced Mason circuit shows that $I_{m2}$ is ideally in reverse of $I_m$ so their zero-crossing points are synchronized at steady-state. The effects of non-idealities can be analyzed in a similar manner. Nevertheless, it is found that shorting the secondary side increases the real part of impedance seen from the primary side and has a negative impact on overall efficiency. This method can be regarded as trading-off efficiency for circuit simplicity.

\subsection{Design Considerations}

\begin{figure}[h]
    \centering
    \includegraphics[width = 0.9\linewidth]{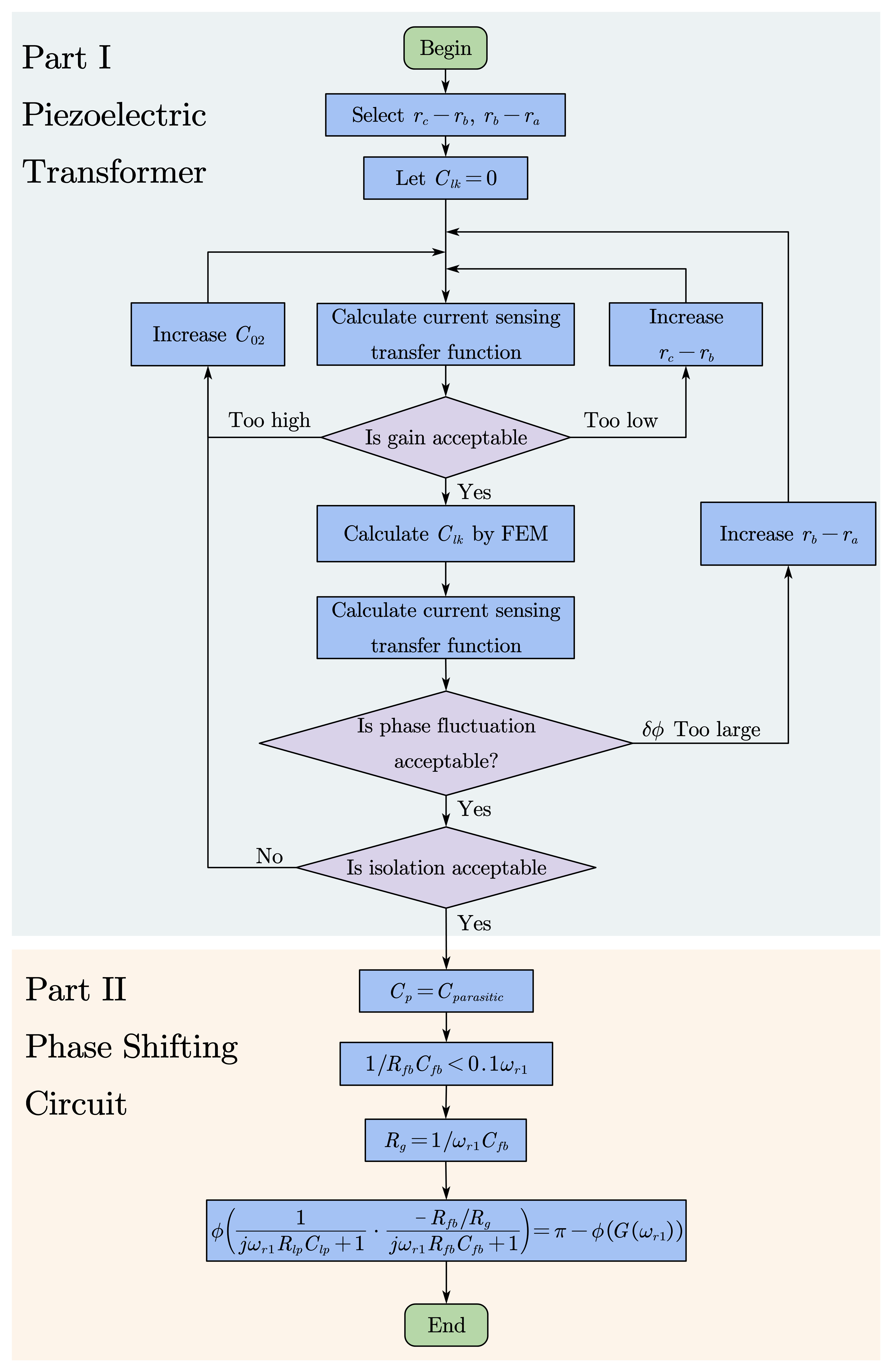}
    \caption{Design flowchart.\label{fig:design-flowchart}}
\end{figure}

The design procedure of the current-sensing PTs aims to realize an acceptable current-sensing transfer function for the desired operating point. First, the impact of the geometric dimensions should be examined. The outer radius $r_c$ and the thickness $h$ are assumed to be pre-determined by the PR optimization process~\cite{bolesEvaluatingPiezoelectricMaterials2022a}, while the electrode radiuses $r_a$ and $r_b$ can be altered to design the sensor ring. To simplify the design procedure, the width of the ring sensor $(r_c-r_b)$ and the isolation middle section $(r_b-r_a)$ are selected as the variables. Fig.~\ref{fig:design-flowchart} depicts the design procedure.

\begin{figure}[h]
    \centering
    \includegraphics[trim=0 8mm 10mm 0, clip, width = 0.9\linewidth]{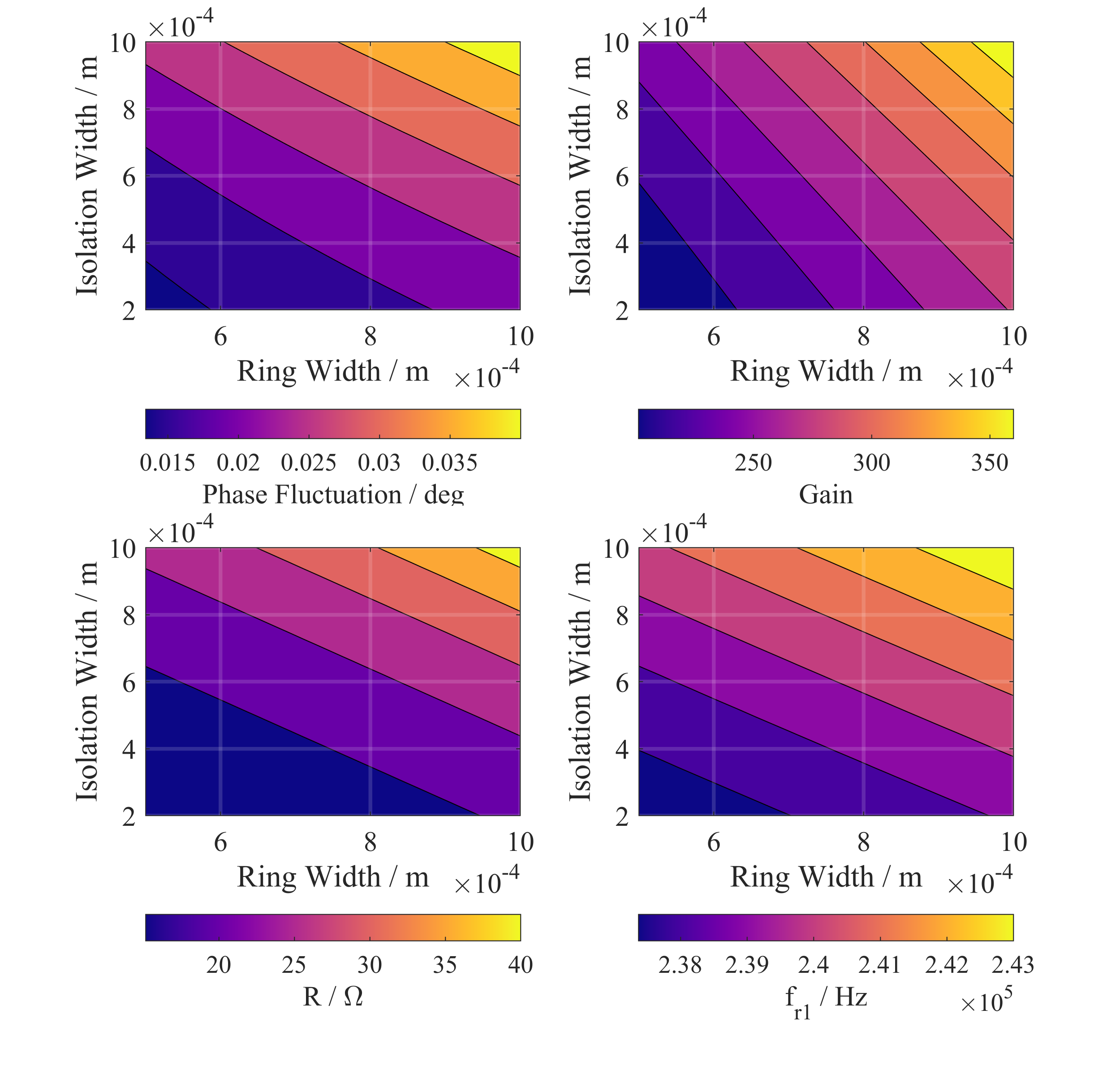}
    \caption{Calculated current-sensing transfer function, equivalent resistance $R$ and resonant frequency $f_{res}$ with varying $r_a$ and $r_b$. $r_c = 5{\rm mm}$, $h = 0.5{\rm mm}$, $\delta_m = 0.5\%$, $\delta_e = 0.5\%$\label{fig:geo-sweep}}
\end{figure}

Fig.~\ref{fig:geo-sweep} maps the impact of the geometries of the sensor ring on the current-sensing transfer function as well as the equivalent resistance $R$ in Fig.~\ref{fig:mason-model} and resonant frequency $f_{r1}$ of the PR. It shows that the phase of the current-sensing transfer function is insensitive to the geometric design, whereas the gain is strongly correlated to the area ratio between the dot and ring sections. Reducing the area of the dot section results in larger equivalent resistance in the motional branch and causes more loss. Besides, the current-sensing transfer function has a considerably high gain, it is thus more desirable to limit the gain to reduce the voltage stress on the downstream circuits.

Both considerations require a smaller sensing ring, so the starting point should be the smallest possible value. If the gain is still to high, then an external capacitor $C_{adj}$ should be placed in parallel with the secondary side to increase $C_{02}$ as shown in Fig.~\ref{fig:phase-shifter}. Otherwise the ring width can be increased for a larger gain. 

Based on similar arguments, the width of the isolation middle section $r_b-r_a$ should be as small as possible. But it also affects the leakage capacitance and consequently the phase fluctuation. The corresponding $C_{lk}$ should be extracted using FEM then substituted into the current-sensing transfer function to examine whether the phase fluctuation between the first resonant frequency and the first anti-resonant frequency is within acceptable range.

Lastly, when criteria of the current-sensing transfer function are met, it should be examined whether the isolation gap $r_b-r_a$ can withstand the maximum voltage between the two electrodes on the same side. For the topology shown in Fig.~\ref{fig:pr-topology}, the secondary side voltage can be found by~\eqref{eq:max-im-vsen}~\cite{bolesEnumerationAnalysisDC2021a}.
\begin{subequations}
    \begin{align}
    \max {I_m} &= \pi \left( {{P_{out}}\frac{{{V_{in}} - {V_{out}}}}{{{V_{in}}}} + f{C_0}{V_{in}}} \right)
    \label{eq:max-im}
    \\
    \max {V_{sen}} &= \max {I_m}\left| {G({\omega _{r1}})} \right|
    \label{eq:max-vsen}
    \end{align}
    \label{eq:max-im-vsen}
\end{subequations}

The design of the phase-shifting circuit is attached as the second part in Fig.~\ref{fig:design-flowchart}. The pole of the integrator should be lower than one tenth of the resonant frequency for a phase-frequency response as flat as possible, and robustness against potential resonant frequency shift due to temperature variation. $R_g$ is set to provide unity gain at the resonant frequency. To simplify the circuit, $C_p$ can be implemented by the parasitic input capacitance of the opamp in Fig.~\ref{fig:phase-shifter}. The value of $R_{in}$ and $R_g$ are used to compensate possible phase error due to non-idealities of the PT and should be determined by $\phi \left( {{I_{m*}}/{V_{sen}}} \right) = \pi  - \phi \left( {G({\omega _{r1}})} \right)$, in which $G(\omega_{r1})$ is obtained from the PT design.

Apart from the aforementioned electrode configuration, in which the center section is used as the power port and the outer section is the sensing port, other designs may also be applicable. E.g. swapping the positions of the power and sensing sections, or utilizing arc-shaped sensor that does not cover the entire circle. Nevertheless, these alternatives are not discussed as they may result in spurious modes due to the break of symmetry or be more sensitive to axial forces applied by the contacts.

\section{Control Strategy}

\hbadness=3000

A PR-based non-isolated topology works within a specific input/output range with a switching sequence, just like the ordinary magnetic-based ones. Nevertheless, unlike the ordinary converters of similar functions, the topologies where PRs are the only energy storage components require a sequence of six or eight stages~\cite{bolesEnumerationAnalysisDC2021a} to realize ZVS, which makes the control strategy equally more complicated. Moreover, some of the stages are constrained by the motional current of PR that is hard to detect. As a result, the existing control strategies rely on sophisticated hardware and software resources. This has been one of the major obstacles to the practical application of PR-based power conversion.

\begin{figure}[h]
    \centering
    \includegraphics[width = 0.9\linewidth]{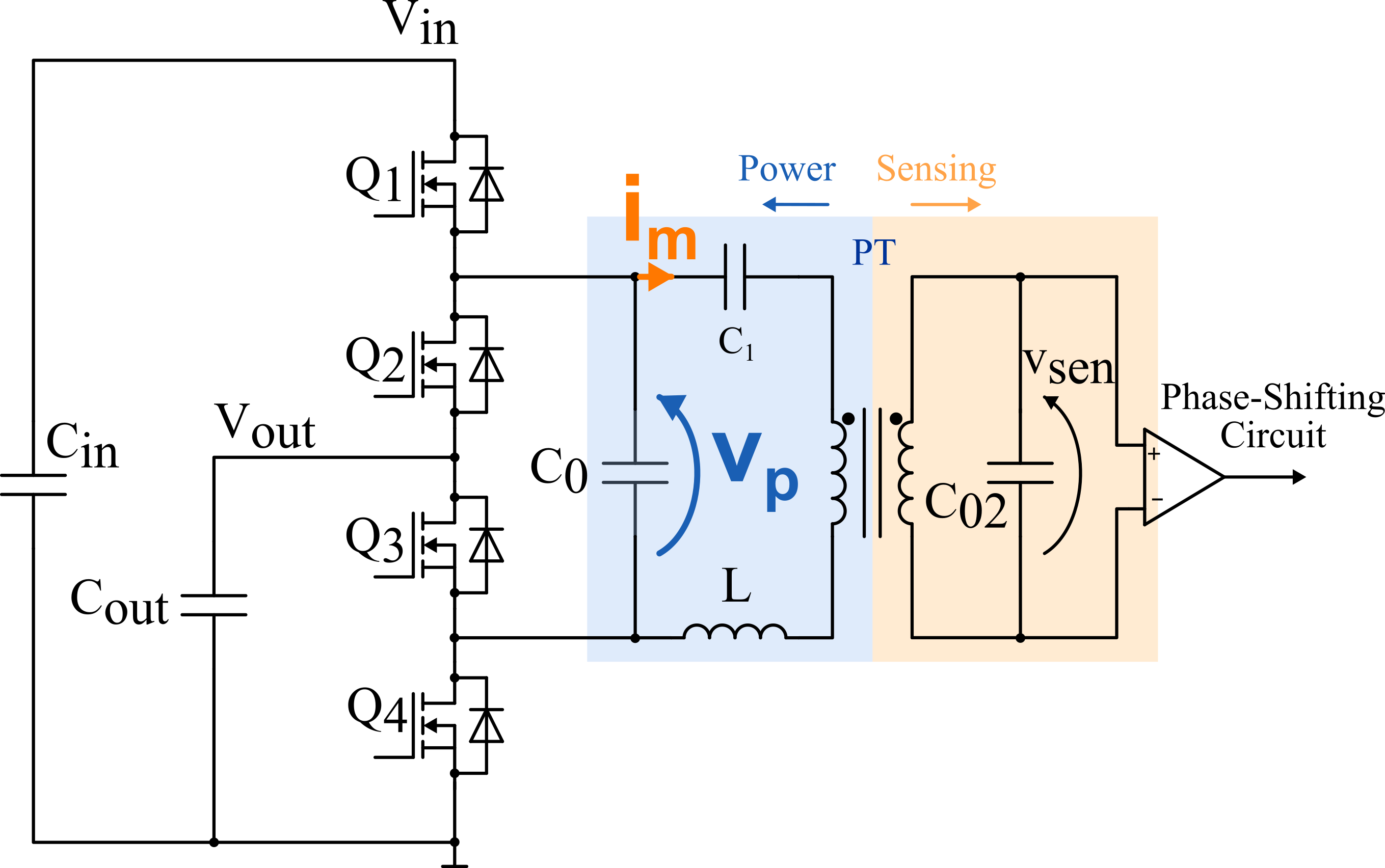}
    \caption{The PR-based step-down converter with the PR replaced by a non-isolated PT.}
    \label{fig:pr-based-converter}
\end{figure}

With the aforementioned current-sensing method, the control strategy can be significantly simplified. It is demonstrated in this work on the step-down topology in Fig.~\ref{fig:pr-topology}, with the PR replaced by a current-sensing PT as shown in Fig.~\ref{fig:pr-based-converter}. $C_0$, $C_1$ and $L$ in the simplified equivalent circuit is still referred to as the PR for simplicity. $C_{02}$ is the secondary side static capacitance. $N$ is the equivalent turns ratio of the ideal transformer. $v_{pa}(t)$ and $v_{pb}(t)$ denote the node voltages of the PR, $v_p(t) = v_{pa} - v_{pb}$. $v_{sen}(t)$ is the sensing voltage.

The switching sequence of the selected topology is listed in Table~\ref{tab:stages}. Although it is categorized as a six-stage sequence, the last stage is split to two in practice for ZVS~\cite{bolesEnumerationAnalysisDC2021a}. To avoid confusion, the sequence is considered to have seven stages in this section.

\begin{table}[h]
    \centering
    \caption{Switching Sequence of the Selected Topology}
        \begin{tabular}{llll}
        \toprule
        Stage & On-Switches & Constraint & Purpose \\
        \midrule
        $S_{12}$ & $Q_1$, $Q_3$ & $T_{12}$ & External Exchange \\
        $S_{23}$ & $Q_3$ & $v_{p3} = 0$ & Soft Switching \\
        $S_{34}$ & $Q_2$, $Q_3$ & $i_{m4} = 0$ & Internal Exchange \\
        $S_{45}$ & $Q_2$ & $v_{p5} = V_{out}$ & Soft Switching \\
        $S_{56}$ & $Q_2$, $Q_4$ & $T_{56}$ & External Exchange \\
        $S_{66B}$ & $Q_4$ & $i_{m6B} = 0$ & Soft Switching \\
        $S_{6B1}$ & $Q_1$ & $v_{p1} = V_{in} - V_{out}$ & Soft Switching \\
        \bottomrule
        \end{tabular}
    \label{tab:stages}
\end{table}

Among the seven stages, the n-th stage $S_{nm}$ is named by their respective starting point $t_n$ and ending point $t_m$. They are categorized into connected stages ($S_{12}$, $S_{34}$, $S_{56}$) and open stages ($S_{23}$, $S_{45}$, $S_{66B}$, $S_{6B1}$). During the former type of stages, the voltage across the primary side of the PT is constrained, and the energy is exchanged between the PT and the external circuits or within the PT itself. As for the latter type, at least one node of the primary side is left floating in order to have $v_p(t)$ resonate to the initial voltage of the next stage, hence to realize ZVS. With the loss factors ignored (i.e. $R=0$), the conservation of energy (CoE) during connected and open stages are given in~\eqref{eq:coe}, and the conservation of charge (CoC) during open stages are given in~\eqref{eq:coc}.

\begin{subequations}
    \begin{equation}
        \left\{ \begin{array}{l}
        {C_1}{\left( {{v_{c[1,3,5]}} - {v_{p[1,3,5]}}} \right)^2} + Li_{m[1,3,5]}^2\\
        = {C_1}{\left( {{v_{c[2,4,6]}} - {v_{p[2,4,6]}}} \right)^2} + Li_{m[2,4,6]}^2
        \end{array} \right.
    \end{equation}
    \begin{equation}
        \left\{ \begin{array}{l}
        {C_0}v_{p[2,4,6,6B]}^2 + {C_1}v_{c[2,4,6,6B]}^2 + Li_{m[2,4,6,6B]}^2\\
        = {C_0}v_{p[3,5,6B,1]}^2 + {C_1}v_{c[3,5,6B,1]}^2 + Li_{m[3,5,6B,1]}^2
        \end{array} \right.
    \end{equation}
    \label{eq:coe}
\end{subequations}
\begin{equation}
    \left\{ \begin{array}{l}
    {C_1}{v_{c[2,4,6,6B]}} + {C_0}{v_{p[2,4,6,6B]}}\\
    = {C_1}{v_c}_{[3,5,6B,1]} + {C_0}{v_{p[3,5,6B,1]}}
    \end{array} \right.
    \label{eq:coc}
\end{equation}

\begin{figure}[h]
    \centering
    \includegraphics[width = 0.9\linewidth]{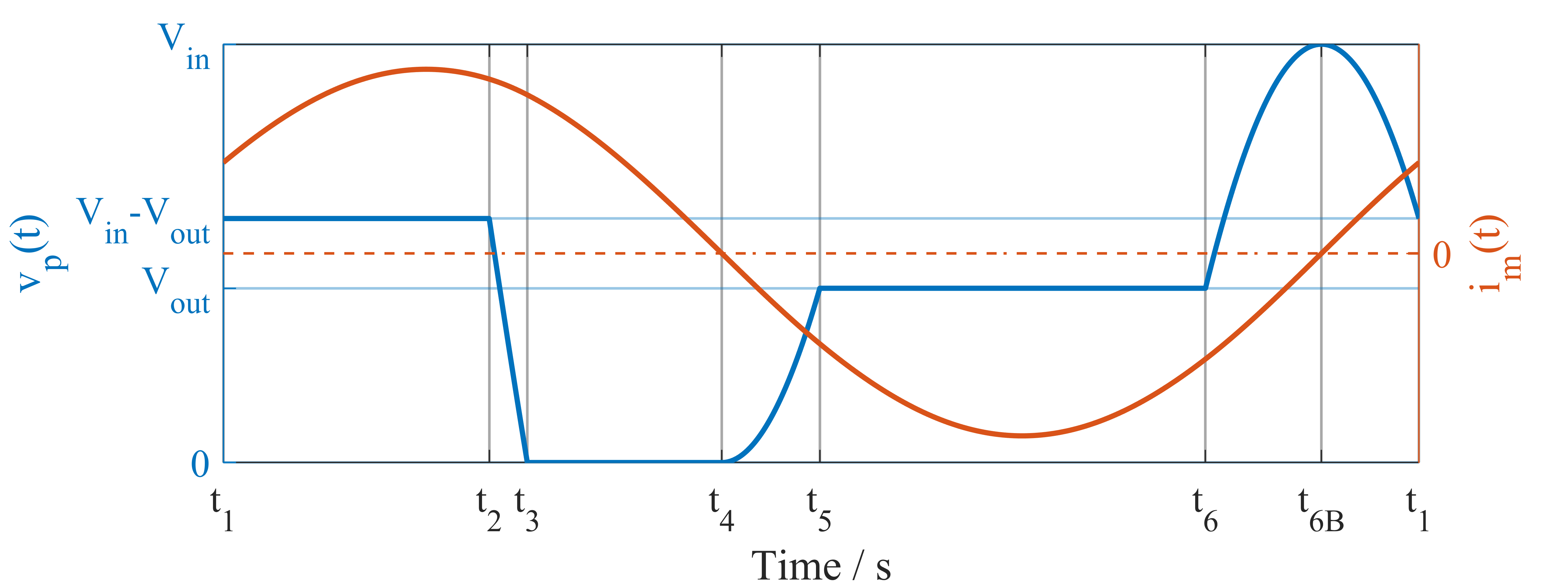}
    \caption{The ideal operating waveform of the primary side voltage $v_p(t)$ and the primary side motional current $i_m(t)$.}
    \label{fig:ideal-waveform}
\end{figure}

In~\eqref{eq:coe} and~\eqref{eq:coc}, the voltages of $C_0$, $C_1$ and the motional current in the equivalent circuit at the beginning of the x-th stage is denoted as $v_{px}$, $v_{cx}$ and $i_{mx}$, respectively. These state variables lead to 21 degrees of freedom in total. $v_{px}$ in the equations are pre-determined as shown in Fig.~\ref{fig:ideal-waveform} to realize ZVS, providing seven constraints. The CoEs and CoCs cancel another 11 degrees of freedom. Among the rest three degrees of freedom, one controls the voltage conversion ratio, and the other two are constrained by the motional current: The current should reach zero at $t_4$ so that ZCS-off is obtained and current circulation is minimized. Another zero-crossing point is at $t_{6B}$ so that $Q_4$ is ZCS-off and $Q_1$ is ZCS-on. The major concern in the control strategy of PR-based converters is to carefully determine the transition points between the adjacent stages to realize the aforementioned constraints and to ensure that the converter operates stably and efficiently.

\begin{figure}[h]
    \centering
    \includegraphics[width = 0.9\linewidth]{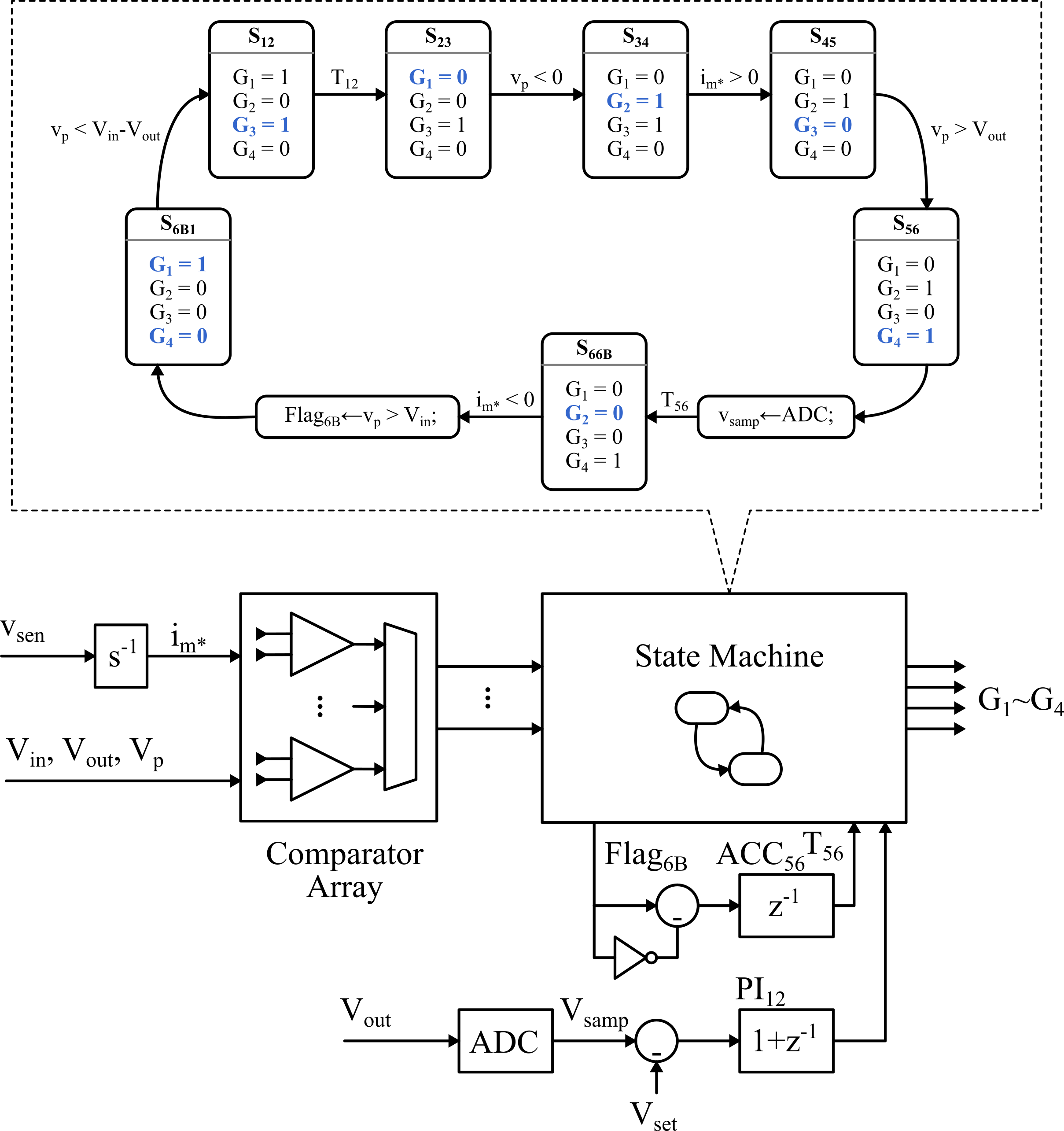}
    \caption{The proposed control system and state machine.\label{fig:control-system}}
\end{figure}

The proposed system in Fig.~\ref{fig:control-system} consists of a state machine, a PI loop, a binary integration loop, five comparators and one low-speed ADC channel. The motional current is extracted from $v_{sen}$ by a phase-shifter and is denoted as $i_{m*}$. The comparators determine five out of seven transitions, while the loops control the other two. The transitions are grouped by their control logic and will be discussed in the rest of this section.

\subsection{Voltage-Triggered Transitions}
During stages $S_{23}$, $S_{45}$ and $S_{6B1}$, $v_p(t)$ is allowed to freely resonate with $v_c$ and $i_m$ until it matches the $v_{px}$ at the beginning of the next stage, hence to achieve ZVS. Given that one side of the PR is tied to certain voltage source during the open stage, whether $v_p(t)$ has reached the ZVS criterion can be easily detected by comparing voltage of the other side and the corresponding reference. This type of transitions are referred to as {\it voltage-triggered transitions}:
\begin{subequations}
    \begin{align}
        {v_{pb}}({t_1}) = {V_{out}} &\Leftrightarrow {v_p}({t_1}) = v_{p1}\\
        {v_{pa}}({t_3}) = {V_{out}} &\Leftrightarrow {v_p}({t_3}) = v_{p3}\\
        {v_{pb}}({t_5}) = 0 &\Leftrightarrow {v_p}({t_5}) = v_{p5}
    \end{align}
\end{subequations}

Being actively triggered every switching cycle, the time span of each stage may vary dramatically, and the state machine may enter deadlock when mis-triggering occurs. This can be avoided by setting upper and lower boundaries for time spans of the stages based on the calculated steady-state solution. From a control perspective, the cycle-to-cycle variation of the control parameters may enhance the stability as compared to some other strategies where they are updated at a lower frequency. This is because the fast response of the system eliminates the latency and avoids right-half-plane zeros (RHPZs).

\subsection{Current-Triggered Transitions}
Although $S_{66B}$ is also an open stage, the transition from $S_{66B}$ to  $S_{6B1}$ should not be voltage-triggered because there are two constraints on $t_{6B}$: $v_p(t_{6B}) = V_{in}$ and $i_m(t_{6B}) = 0$. Between them, the current criterion can always be met, but $v_p(t)$ may not ever reach $V_{in}$ if the energy stored in the piezoceramic is not enough at $t_{6}$, which is controlled by the former stage. Without the current information, this leads to an extra control loop~\cite{pielFeedbackControlPiezoelectricResonatorBased2021a} or PLL~\cite{liuClosedloopControlDualside2025a} and may require high-speed ADC sampling~\cite{liuClosedloopControlDualside2025a} to judge whether $S_{66B}$ ends at the current zero-crossing point and whether $i_m(t_6)$ allows $v_p(t)$ to precisely resonate to $V_{in}$ at $t_{6B}$.

With the motional-current-sensing PT proposed in this work, the motional current can be restored from $v_{sense}$ by a phase-shifter. Then the current zero-crossing point can be accurately located and used to drive the transition at $t_{6B}$. Also, by detecting whether the voltage reaches $V_{in}$ at $t_{6B}$ with another comparator, it can be easily determine that whether $i_{m6}$ is high enough and be used to control the former stage $S_{56}$. The details will be explained later.

Similarly, the other zero-crossing point of $i_m(t)$ is used to determine the end point of the short-circuit stage $S_{34}$, which requires precise voltage sampling right after $t_4$ and another control loop without current information. In conclusion, the restored motional current trigger two of the transition points:
\begin{subequations}
    \begin{align}
        {i_{m*}}({t_{6B}}) & = 0\\
        {i_{m*}}({t_4}) & = 0
    \end{align}
\end{subequations}

\subsection{Loop-Driven Transitions}
Both $S_{12}$ and $S_{56}$ are connected stages, during which $v_p(t)$ stays the same, while $i_{m2}$ and $i_{m6}$ are not constraint to specific reference either. Therefore, it is impossible to control these stages by comparator-triggered events. Instead, their time span should be controlled by separate feedback loops with different error inputs. 

\subsubsection{Voltage Regulation Loop} 
As mentioned earlier, $T_{12}$ is used to regulate the voltage conversion ratio, this can also be verified by CoEs and CoCs:
\begin{align}
    \left( {{V_{in}} - {V_{out}}} \right){q_{12}} + 0 \cdot {q_{34}} + {V_{out}}{q_{56}} = 0
    \nonumber
    \\
    \frac{{{V_{out}}}}{{{V_{in}}}} = \frac{{{q_{12}}}}{{{q_{12}} - {q_{56}}}} = \frac{{{q_{12}}}}{{2{q_{12}} + {q_{34}}}}
\end{align}

Observed from Fig.~\ref{fig:ideal-waveform}, the increase of $T_{12}$ enlarges $q_{12}$ and reduces $q_{34}$, making $T_{12}$ positively correlated to the voltage conversion ratio. Therefore, the error between the sampled output voltage and the reference is fed to a digital PI controller that manipulates $T_{12}$. Initial output and saturation limit of the controller are set based on the numerical solution of the constraints to ensure stability. 

The ADC can either directly samples $V_{out}$ or samples the primary side voltage of the PT, $v_p$. The former approach is more desirable as it allows the sampling rate of the ADC to be near the resonant frequency or even lower, if the transient response time is not of great concern. Because of the continuity of $V_{out}$, it is also possible to remove the ADC and implement the PI controller on analog devices. Alternatively, when connected to $v_p$ instead, the ADC samples the output voltage during $S_{56}$ and latches the data. During the rest of the switching cycle, the ADC can track on the $v_p$ waveform to provide $V_{in}$ information and detect fault waveform at the expense of higher sampling rate. For simplicity, the prototype in Section IV implements the former configuration.

\subsubsection{ZVS Loop with Binary Feedback}

\begin{figure}[h]
    \centering
    \includegraphics[trim=15mm 0 15mm 0, clip, width = \linewidth]{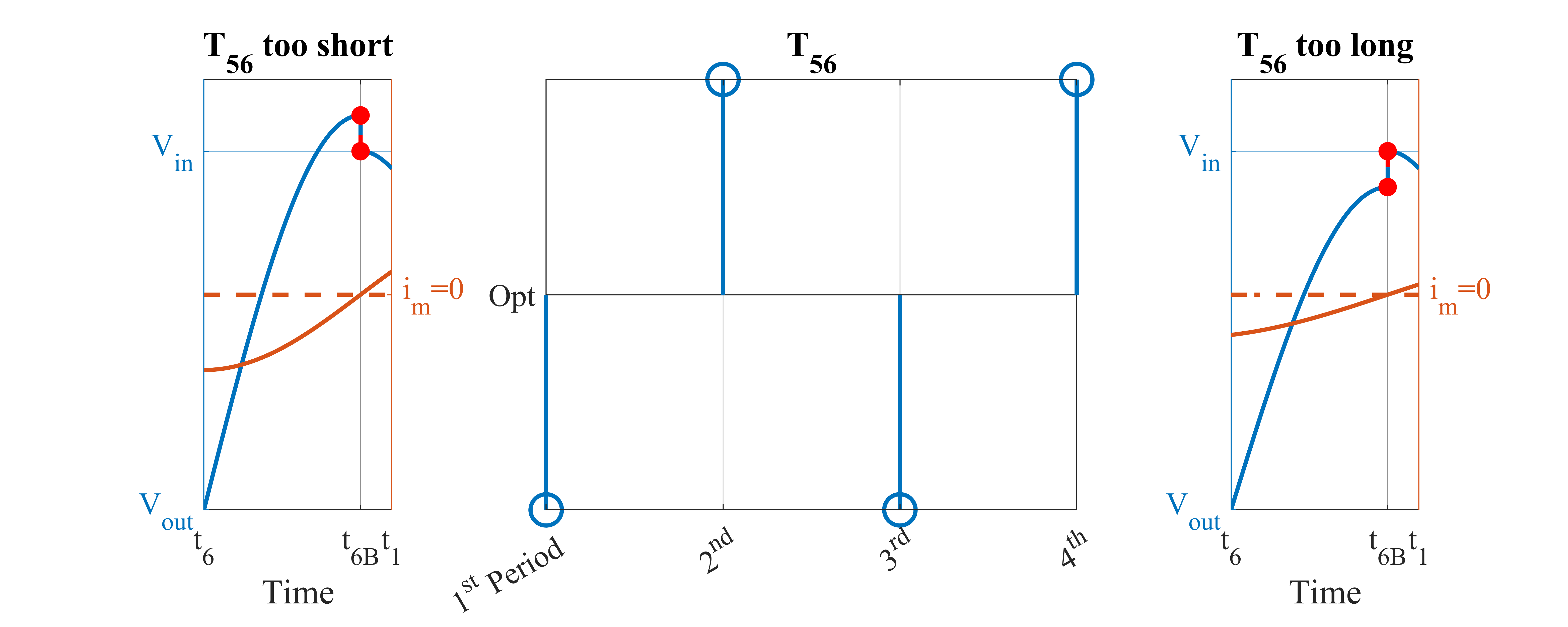}
    \caption{Output of the ZVS loop at equilibrium.}
    \label{fig:pi6b-output}
\end{figure}

Another loop is implemented to control the current at the end of stage $S_{56}$ so that $v_p$ resonates exactly to $V_{in}$ at $t_{6B}$. Instead of taking the exact voltage values read from ADC as the input~\cite{liuClosedloopControlDualside2025a}, this loop utilizes the binary output of a comparator. A similar method has been adopted in~\cite{touhamiImplementationControlStrategy2020}. When $S_{66B}$ transits to $S_{6B1}$ at $t_{6B}$, the controller latches the comparator readout of whether $v_p > V_{in}$. If so, the signal $Flag_{6B}$ is set to 1, so the next $T_{56}$ increases, reducing the initial current of stage $S_{6B}$ and therefore reducing $v_p(t_{6B})$, vice versa. Because of the binary input, the ZVS loop has no steady state and the output will oscillate around the optimal $T_{56}$ as illustrated in Fig.~\ref{fig:pi6b-output}. But the oscillation can be limited by choosing an appropriate $K_a$ in~\eqref{eq:acc56}.

\begin{equation}
    {T_{56}}(k + 1) = {T_{56}}(k) + {K_a}\left( {Fla{g_{6B}} - \overline {Fla{g_{6B}}} } \right)
    \label{eq:acc56}
\end{equation}

The required hardware and software resources of the control strategies proposed in this work and the others are compared in Table~\ref{tab:comparison}. By shifting the burden from software and high-speed ADCs to comparators and opamps, this work reduces the system cost and expand the possible operating frequency. The prototype is controlled by FPGA only for convenience in experiments, while it is also possible to build the system on cost-effective MCUs, DSPs or CPLDs because the state machine and the two digital loops occupy little resources. Besides, the simple structure makes it easier to implement the control system on customized ICs.

\begin{table}[h]
    \centering
    \caption{Resources Comparison of Closed-Loop Control Strategies}
        \begin{tabular}{llrrrr}
        \toprule
            & RTC$^\diamond$ & PID & PLL & ADC & Comparators\\
        \midrule
        \cite{pielFeedbackControlPiezoelectricResonatorBased2021a} & $-$   & 5     & 0     & 3$^\circ$ & 0  \\
        \cite{forresterBidirectionalInvertingPiezo2023a} & $-$  & 1 & 1 & 1     & 3 \\
        \cite{bigotNewClosedloopRegulation2024} & $+$ & 1 & 0 & 1 & 0 \\
        \cite{liuClosedloopControlDualside2025a} & $-$   & 2     & 1     & 3$^\circ$ & 2 \\
        \cite{stoltCurrentModeControl2025a} & $+$   & 2     & 0     & 2     & 0 \\
        \cite{touhamiClosedLoopControlSymmetric2026} & $-$ & 4 & 0 & 3$^\circ$ & 0 \\
        This  & $-$  & 1     & 0     & 1     & 5 \\
        \bottomrule
        \end{tabular}
    \label{tab:comparison}
    \\
    \vspace{0.1cm}
    $\circ$ : High-speed ADC / short sampling windows; \\
    $\diamond$ : Real-time calculation of switching timing.    
    
\end{table}

\section{Implementation and Experiment}

\subsection{Ring-Dot PT}

\begin{figure}[h]
    \centering
    \includegraphics[width = 0.7\linewidth]{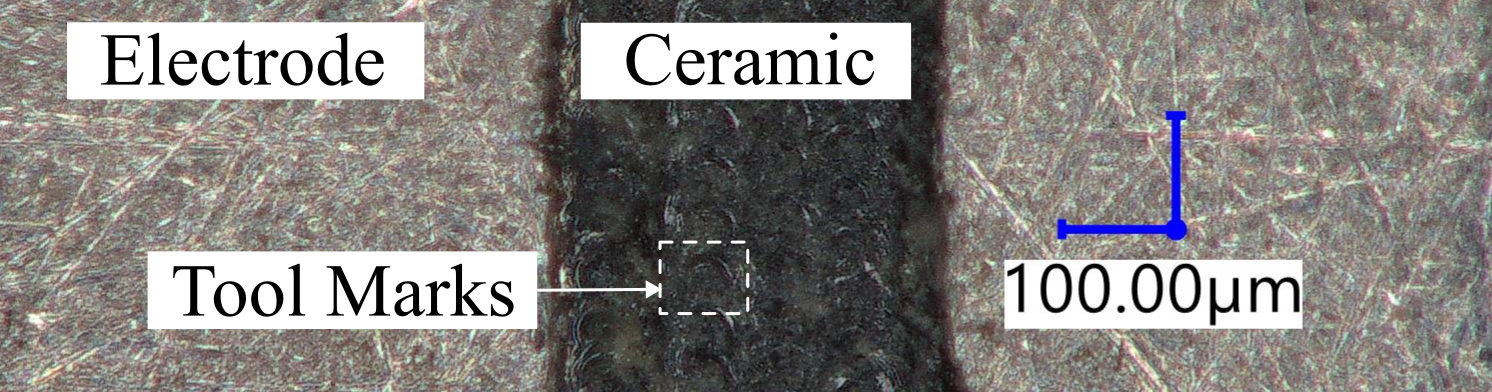}
    \caption{Middle annular section of PT-I after electrode removal.\label{fig:pt-middle-section}}
\end{figure}

Two ring-dot PTs are fabricated for experiment based on piezoceramic item 2006 and 2470 provided by APC International. Both surfaces of the components were fully covered with silver electrodes. The dot and ring sections are separated by using a CNC milling machine to mechanically remove the annular electrodes on both sides corresponding to the middle ring section (Fig.~\ref{fig:pt-middle-section}). During the removal process, the engraving depth was gradually increased to minimize the damage on the piezoceramic. The final depth is measured under digital microscope to be $\approx  60 {\rm \mu m}$ for the PT with a thickness of 2.7mm, and $\approx 30 {\rm \mu m}$ for the 0.5mm one. Properties of the fabricated components are measured and listed in Table~\ref{tab:pt-property}.

Although it was mentioned in Section II that smaller sensing rings are favorable, the width of the ring sections is limited by the mechanical fabrication process. A photolithography-based process flow~\cite{navalHighEfficiencyIsolatedPiezoelectric2026} can be applied as a solution.

\begin{figure}[h]
    \centering
    \includegraphics[width = 0.7\linewidth]{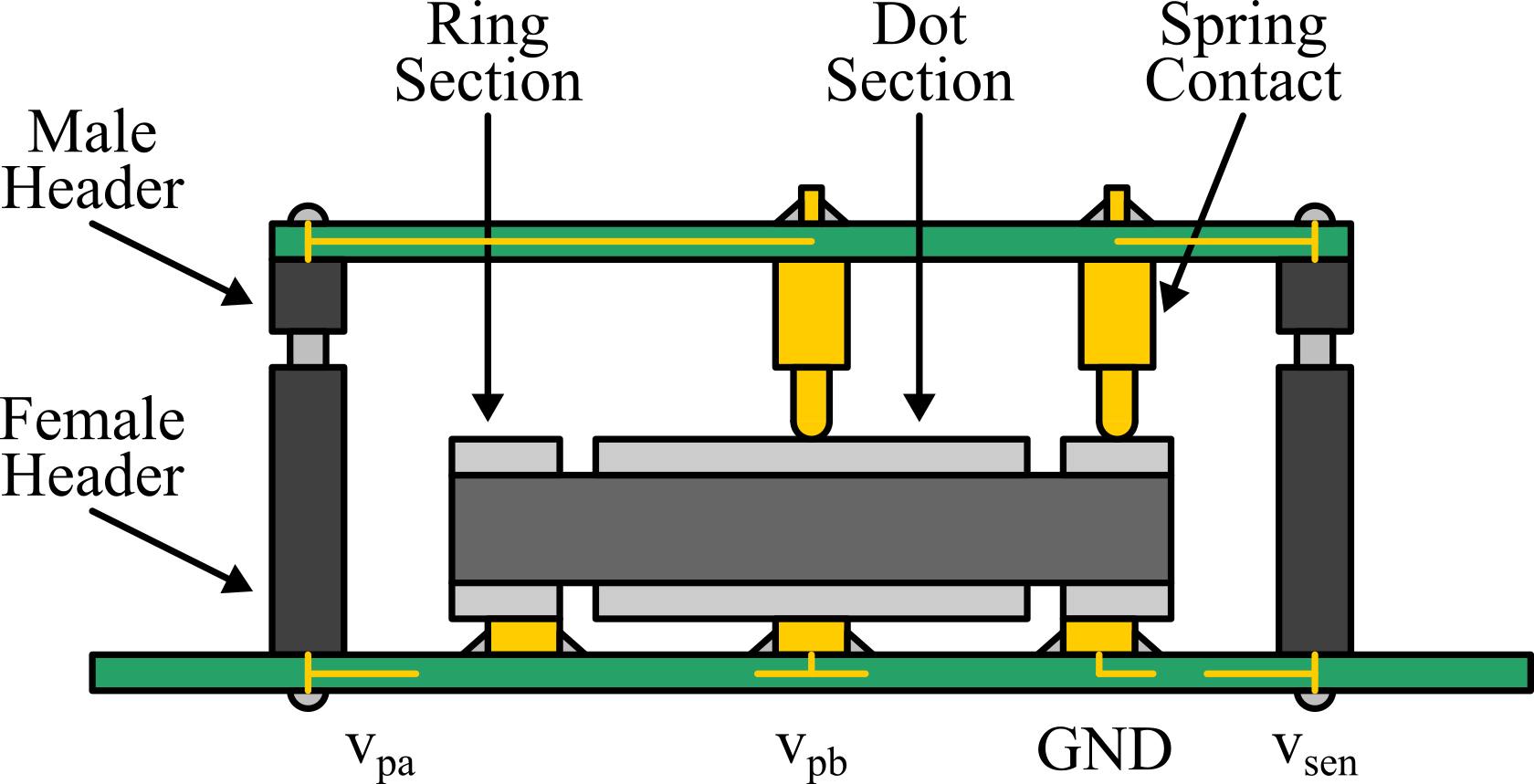}
    \caption{Mounting of the ring-dot PT on the prototype.}
    \label{fig:mounting}
\end{figure}

In this work, a daughter board with spring-loaded connectors (pogo pins) mounts the PTs to the main board. The two boards are connected via 2.54mm pin headers and are secured by friction provided by the socket. The insertion depth of the daughter board was manually adjusted during experiment to make the spring connectors just touch the surface electrodes without significantly disrupting the vibration of the component. The mounting scheme is illustrated in Fig.~\ref{fig:mounting}.

\begin{figure}[h]
    \centering
    \includegraphics[trim=0 0 0 5mm, clip, width = 0.7\linewidth]{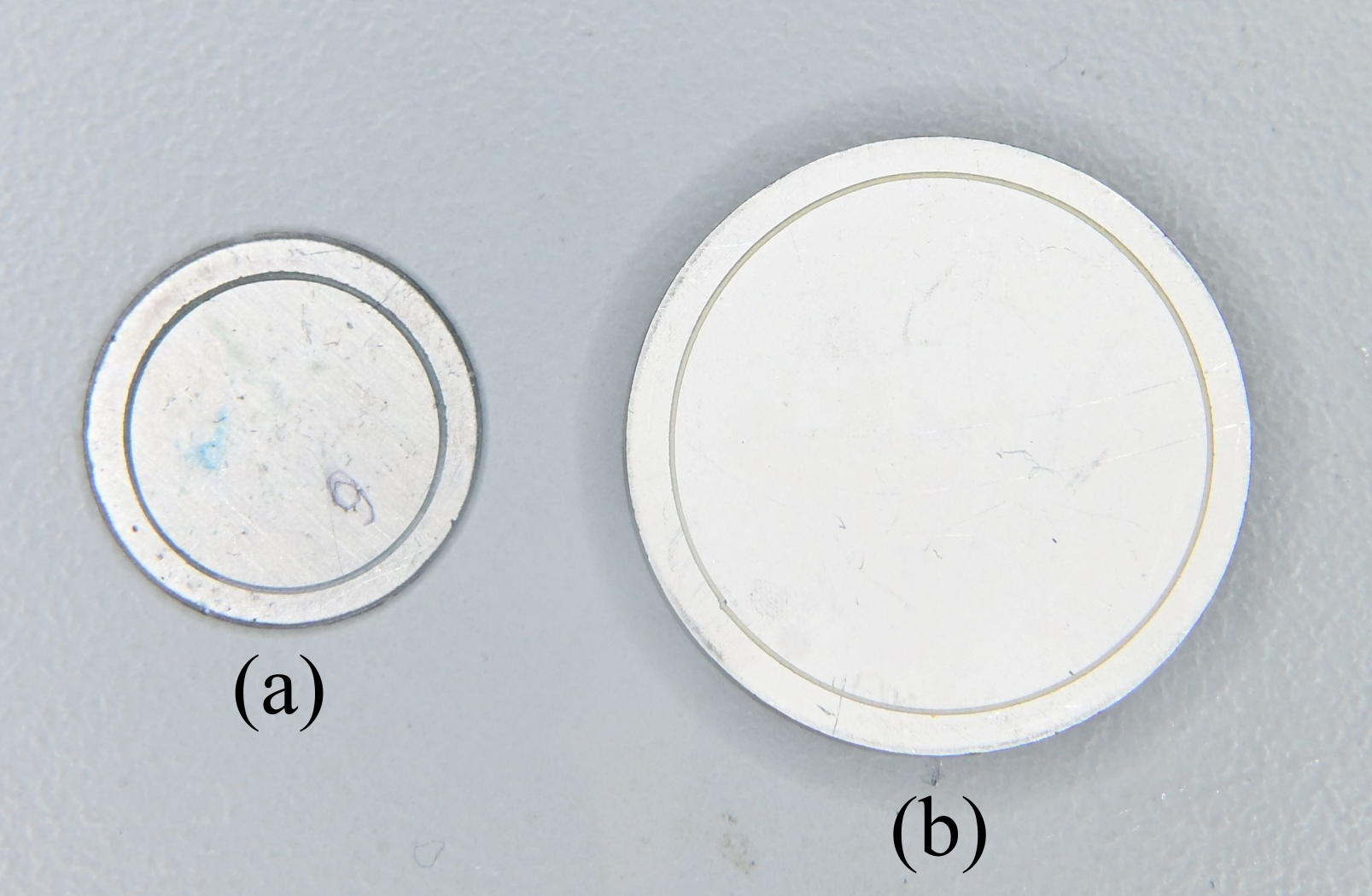}
    \caption{Manufactured PTs. (a) PT-I. (b) PT-II.\label{fig:pt-photograph}}
\end{figure}

\begin{table}[h]
  \centering
  \caption{PT Properties}
    \begin{tabular}{llllll}
    \toprule
          & PT-I  & PT-II  &       & PT-I  & PT-II \\
    \midrule
    Material & APC 880 & APC 841 & $f_r1$ & 156.4 kHz & 98.2 kHz \\
    $r_a$ & 6.23 mm & 9.98 mm & $C_0$ & 1.8 nF & 1.1 nF \\
    $r_b$ & 6.40 mm & 10.09 mm & $C_1$ & 463.4 pF & 294.9 pF \\
    $r_c$ & 7.25 mm & 11.50 mm & $L$   & 2.7 mH & 8.9 mH \\
    $h$   & 0.5 mm & 2.7 mm & $R$   & 6.8 $\Omega$ & 3.6 $\Omega$ \\
    $h_{engrave}$ & $\sim 30 \,{\rm \mu m}$& $\sim 60 \,{\rm \mu m}$ & $C_{02}$ & 1.05 nF & 529.3 pF \\
    &&& $N$   & 21.4 & 87.0 \\
    \bottomrule
    \end{tabular}%
  \label{tab:pt-property}%
\end{table}%

\begin{figure}[h]
    \centering
    \includegraphics[trim=10mm 0 10mm 5mm, clip, width = \linewidth]{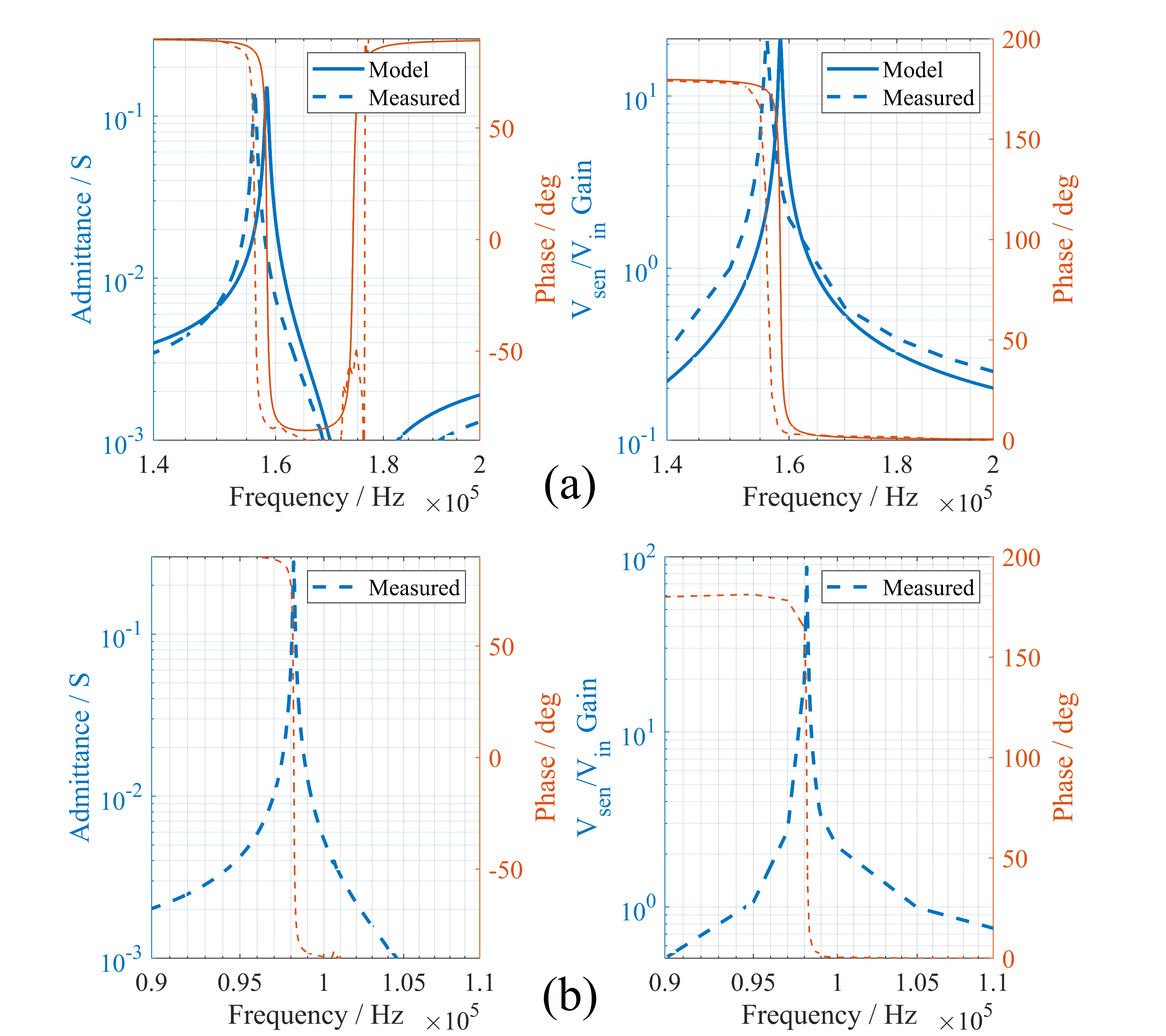}
    \caption{Primary side admittance and primary to secondary side voltage gain of (a) PT-I; (b) PT-II.\label{fig:model-measure}}
\end{figure}

The primary side admittance of PT-I is measured by Vector Network Analyzer (VNA) and the primary to secondary side voltage gain is measured by signal generator and oscilloscope. The data is plotted in Fig.~\ref{fig:model-measure}. For PT-I, the model-calculated results are shown in comparison. Material properties of PZT-8 (Navy Type III) are used in calculation, as claimed by the vendor to be equivalent to APC 880. The dielectric loss factor $\delta_e$ is set to be 0.4\% as claimed, and the mechanical loss factor $\delta_m$ is determined to be 0.3\% in experiment by fitting the peak admittance. It can be observed that apart from a +1.3\% error in resonant frequency and a -1.2\% error in anti-resonant frequency, the agreement between the modelled and measured data is acceptable. The errors can be attributed to the minor difference between the nominal and actual material parameters, e.g. a 2.6\% error in mass density or 1.3\% error in elastic compliance constants.

Although the $V_{sen}/I_{m}$ transfer function cannot be physically measured, experiment results verify the model and indicate the validity of the current-sensing transfer function. The measured data of PT-II is not compared with model estimation due to the lack of material properties.

\subsection{Circuit Implementation}

\begin{figure}[h]
    \centering
    \includegraphics[width = 0.8\linewidth]{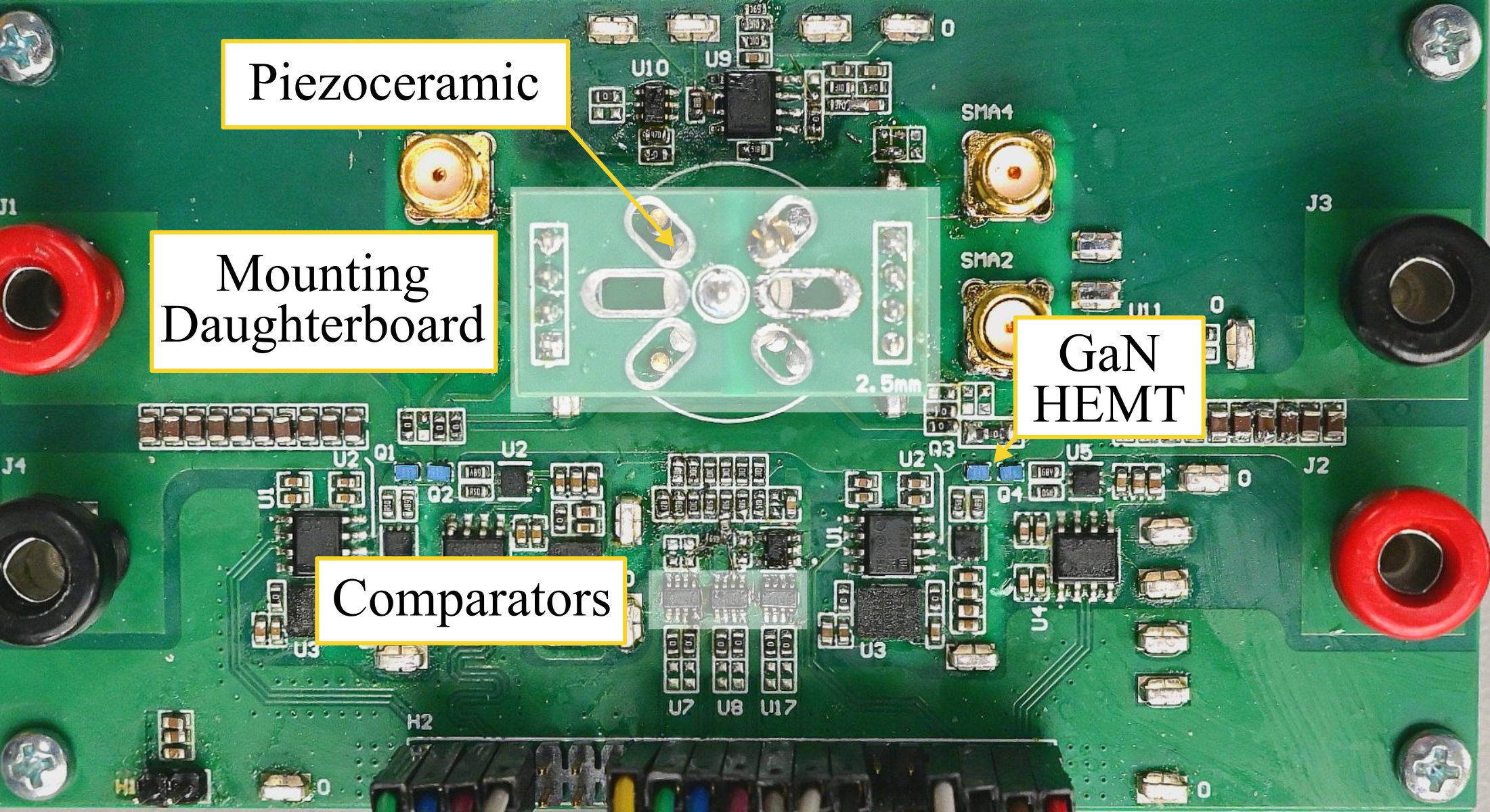}
    \caption{Prototype PCB.\label{fig:pcb}}
\end{figure}

The whole system other than the controller is built on a six-layer PCB shown in Fig.~\ref{fig:pcb}. For flexibility in experiment, the state machine in Fig.~\ref{fig:control-system} is implemented on a Xilinx Zynq 7020 FPGA running at 100 MHz. The main PCB connects to the FPGA development board via jumper wires. Other hardwares in the control system are listed in Table~\ref{tab:part-selection}. 

\begin{table}[H]
  \centering
  \caption{Part Selection}
    \begin{tabular}{ll}
    \toprule
    Component & Part \\
    \midrule
    Comparator & Texas Instruments TLV3502 \\
    Opamp & Texas Instruments TLV3542 \\
    ADC   & Analog Devices AD9215 \\
    Digital Isolator & Skyworks Si8240 \\
    Gate Driver & Texas Instruments LMG1025 \\
    Switching Device & EPC EPC2007C GaN HEMT \\
    \bottomrule
    \end{tabular}
  \label{tab:part-selection}
\end{table}

\begin{table}[H]
    \centering
    \caption{Propagation Delays\label{tab:propagation-delay}}
    \begin{tabular}{ll}
        \toprule
        Component & Propagation Delay \\
        \midrule
        Comparator TLV350x & 4.5 ns \\
        Controller (FPGA for experiment) & $\sim$10 ns \\
        Signal Isolator Si8420Bx & 6.0 ns \\
        Gate Driver LMG1025 & 2.9 ns \\
        \bottomrule
    \end{tabular}
\end{table}

Five comparator cells are used to generate voltage and current events. For each voltage or current triggered event, the feedback chain contains a comparator, a controller, a gate driver and its (integrated or discrete) signal isolator. The propagation delays within the chain are listed in Table~\ref{tab:propagation-delay}. When the resonant frequency or the output power is high, the propagation delay of the control system may be significant and cause the switching device reverse-conduction to enter reverse-conduction before turning on, consequently results in imperfect ZVS. This can be compensated by adjusting the threshold of the comparators so that they can be triggered earlier.

The motional current restoration circuit is designed to be a high input impedance phase-shifting circuit as shown in Fig.~\ref{fig:phase-shifter}. As discussed in Section II-D, different sets of parameters should be designed for different PTs, and the calculated values for PT samples in this work are listed in Table~\ref{tab:phase-shifter}. The input capacitance of the Opamp works as the capacitor in the low-pass filter, $C_{lp}$, to simplify the circuit.

\vspace{-0.5cm}
\begin{table}[H]
  \centering
  \caption{Selected Values of the Phase Shifting Circuit}
    \begin{tabular}{lllllll}
    \toprule
    & $C_{adj}$ & $R_{lp}$  & $C_{lp}^*$ & $R_{g}$ & $R_{fb}$ & $C_{fb}$ \\
    \midrule
    PT-I & 15 nF & $5.1\,{\rm k\Omega}$ & $\sim$3 pF & $1.5\,{\rm k\Omega}$ & $56\,{\rm k\Omega}$ & 680 pF \\
    PT-II & 15 nF & $0\,{\rm \Omega}$ & $\sim$3 pF & $1.5\,{\rm k\Omega}$ & $30\,{\rm k\Omega}$ & 1 nF \\
    \bottomrule
    \end{tabular}
  \label{tab:phase-shifter}
  \\
  \vspace{0.1cm}
  *: Input capacitance of TLV3542.
\end{table}

\subsection{Steady-State Operation}

\begin{figure}[h]
    \centering
    \includegraphics[width = 0.8\linewidth]{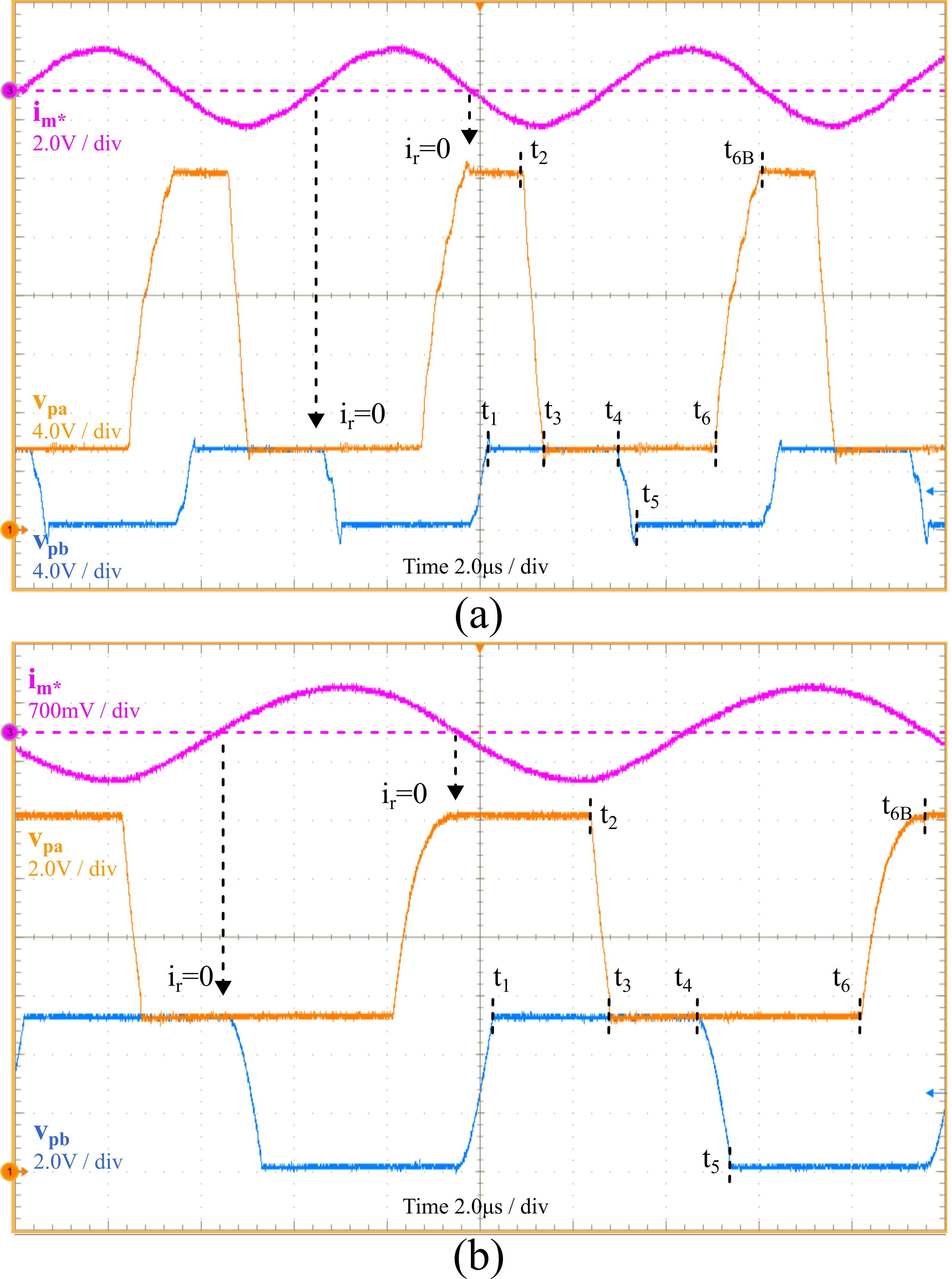}
    \caption{Steady-state operation waveform of converter with (a) PT-I, $V_{in} = 24{\rm V}$, $V_{out} = 5{\rm V}$, $P_{out} = 0.16{\rm W}$; (b) PT-II, $V_{in} = 12{\rm V}$, $V_{out} = 5{\rm V}$, $P_{out} = 0.08{\rm W}$\label{fig:operation-waveform}}
\end{figure}

The acquired steady-state operating waveform of DC/DC down-conversion with both PTs at different operation points are shown in Fig.~\ref{fig:operation-waveform}. Firstly, the $v_{pa}$ and $v_{pb}$ waveforms verify that soft-switching is achieved on all transitions during a switching cycle. Besides, the current-driven transitions achieve ZCS at $t_4$ and $t_{6B}$. Just before $t_{6B}$, $i_{m*}$ falls below the threshold while $v_{pa}$ peaks simultaneously, indicating that the restored motional current $i_{m*}$ correctly reflects the original phase. Meanwhile at $t_{6B}$, the peak of $v_{pa}$ is around $V_{in}$ and slightly oscillates over adjacent switching cycles. This verifies the binary error accumulator that controls $T_{56}$. The other loop-controlled quantity $T_{12}$ is also stabilized and maintains the output voltage around 5V.

\begin{figure}[h]
    \centering
    \includegraphics[width = 0.8\linewidth]{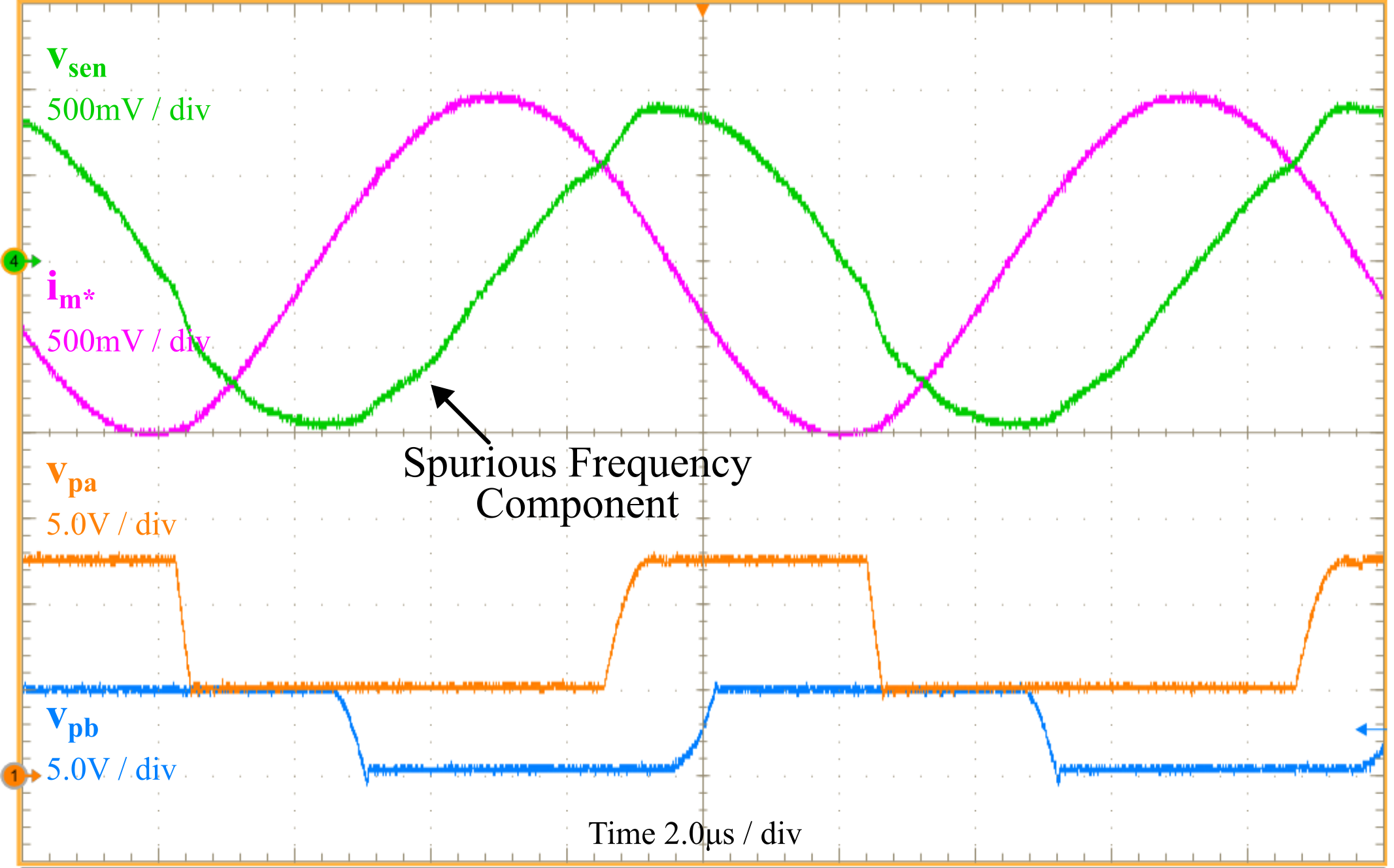}
    \caption{Sensing voltage and restored motional current at steady state with PT-II. $V_{in} = 12{\rm V}$, $V_{out} = 5{\rm V}$, $P_{out} = 0.16{\rm W}$.\label{fig:sensing-voltage}}
\end{figure}

Although multiple spurious resonant frequency components exist in the PR voltage in Fig.~\ref{fig:operation-waveform}a, and consequently in the sensing voltage in Fig.~\ref{fig:sensing-voltage}. The intrinsic low-pass feature of the phase-shifting circuit largely attenuates the noise. The spurious resonance in $v_{pa}$ and $v_{pb}$ may result in unintended or missed triggering of $Flag_{6B}$ in the binary feedback loop. But the binary accumulator also has a strong filtering effect on high frequency signals. Thus the spurious vibration has little impact on ZVS and switching loss.

\begin{figure}[h]
    \centering
    \includegraphics[width = 0.85\linewidth]{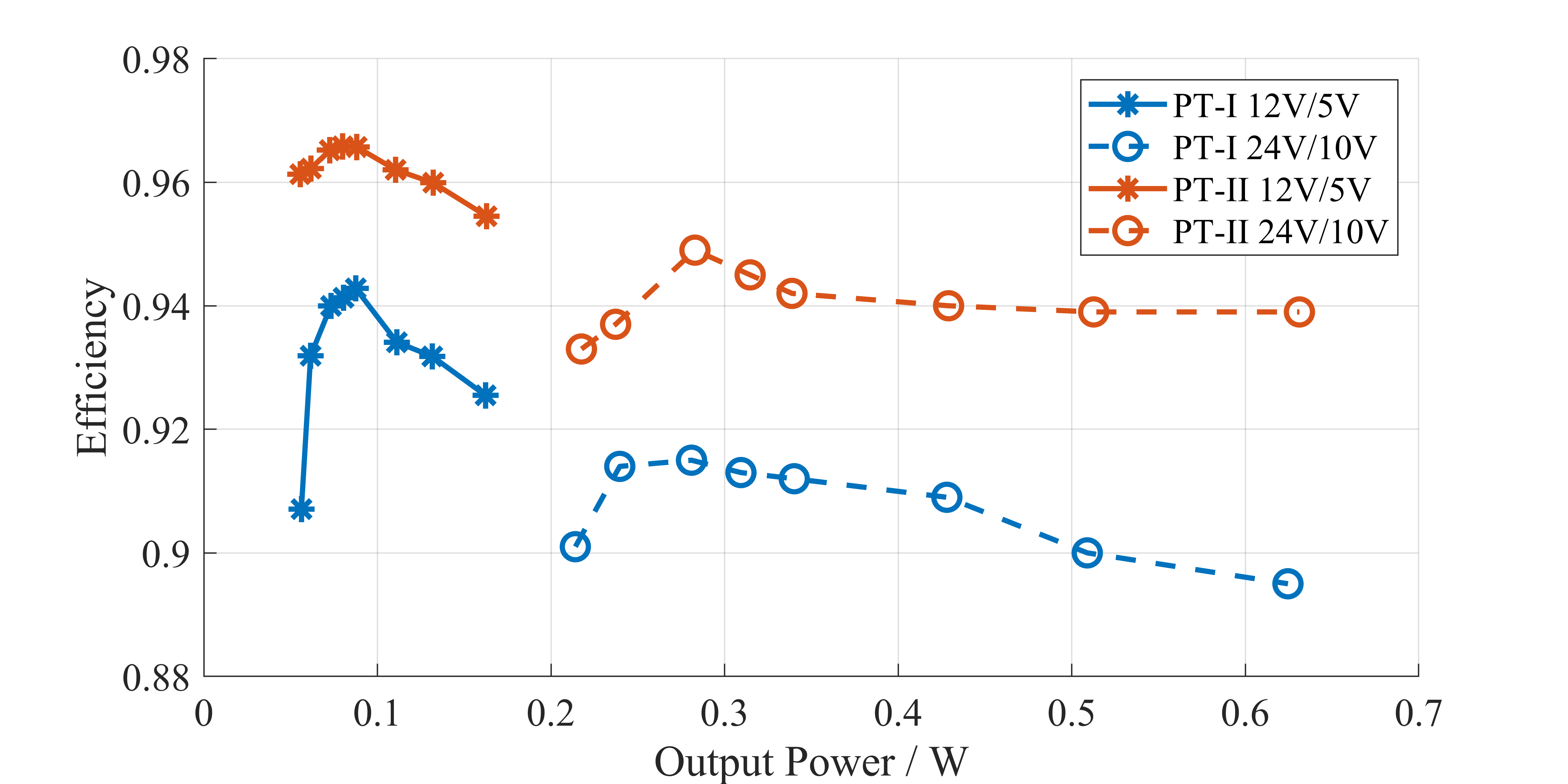}
    \caption{Measured converter efficiency. Auxiliary power consumption is not included.\label{fig:efficiency}}
\end{figure}

The efficiency of the prototype is measured and plotted in Fig.~\ref{fig:efficiency}. The data is calculated by $P_{out}/(V_{in}I_{in})$, without the auxiliary power consumption (FPGA, gate drivers, comparators, etc.). The output power of maximum efficiency will be higher as the input voltage increases~\cite{bolesEnumerationAnalysisDC2021a}. But it is not a major concern of this work. The peak efficiency reaches $94.3\%$ for PT-I and $96.6\%$ for PT-II. It is found during experiment that the efficiency is highly sensitive to the mechanical load applied to the PTs, especially for PT-I that is smaller and thinner.

\begin{figure}[h]
    \centering
    \includegraphics[width = 0.8\linewidth]{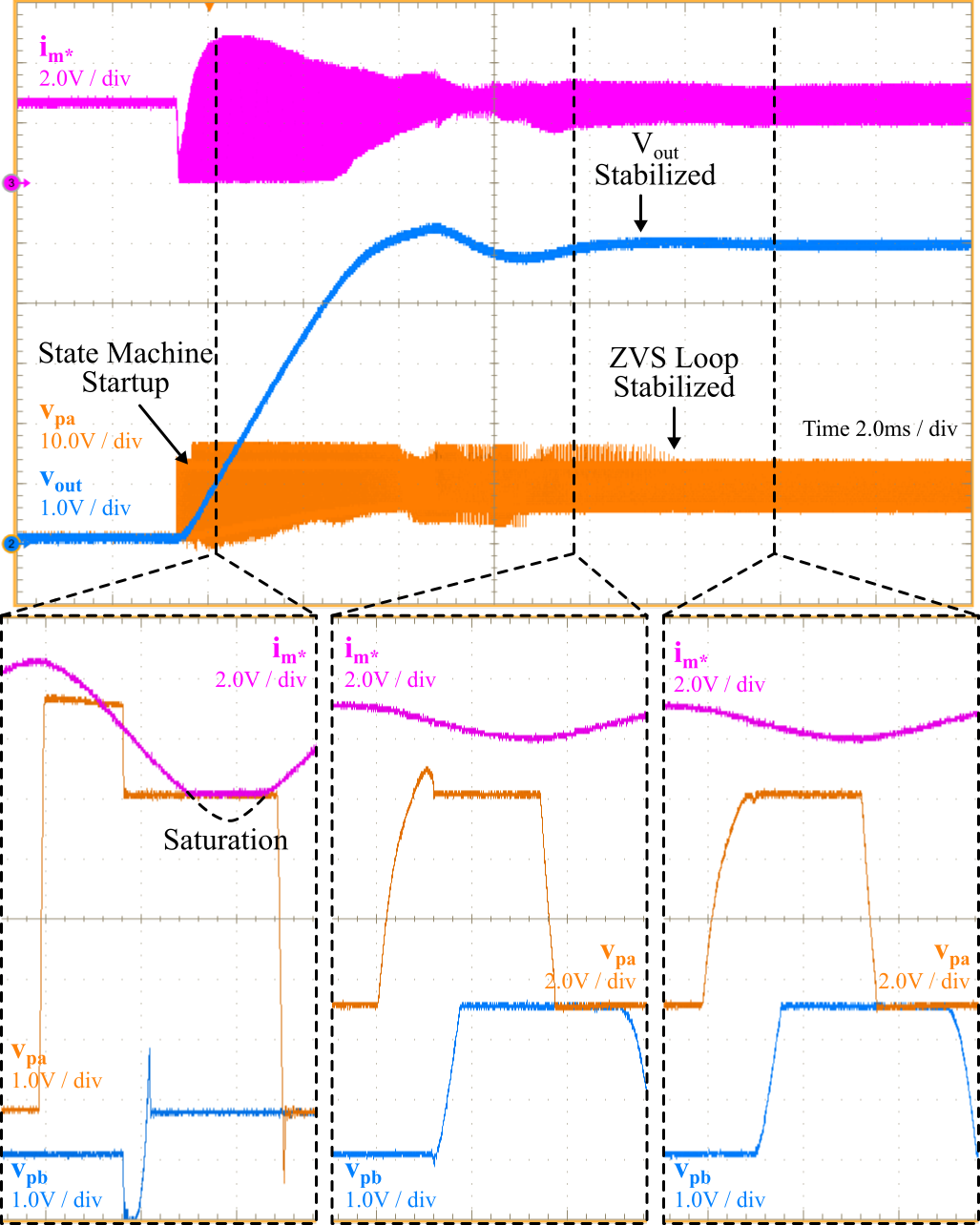}
    \caption{Start-up waveform of converter with PT-II. $V_{in} = 12{\rm V}$, $V_{out} = 5{\rm V}$, $R_{load} = 324\Omega$.\label{fig:startup}}
\end{figure}

\begin{figure}[h]
    \centering
    \includegraphics[width = 0.8\linewidth]{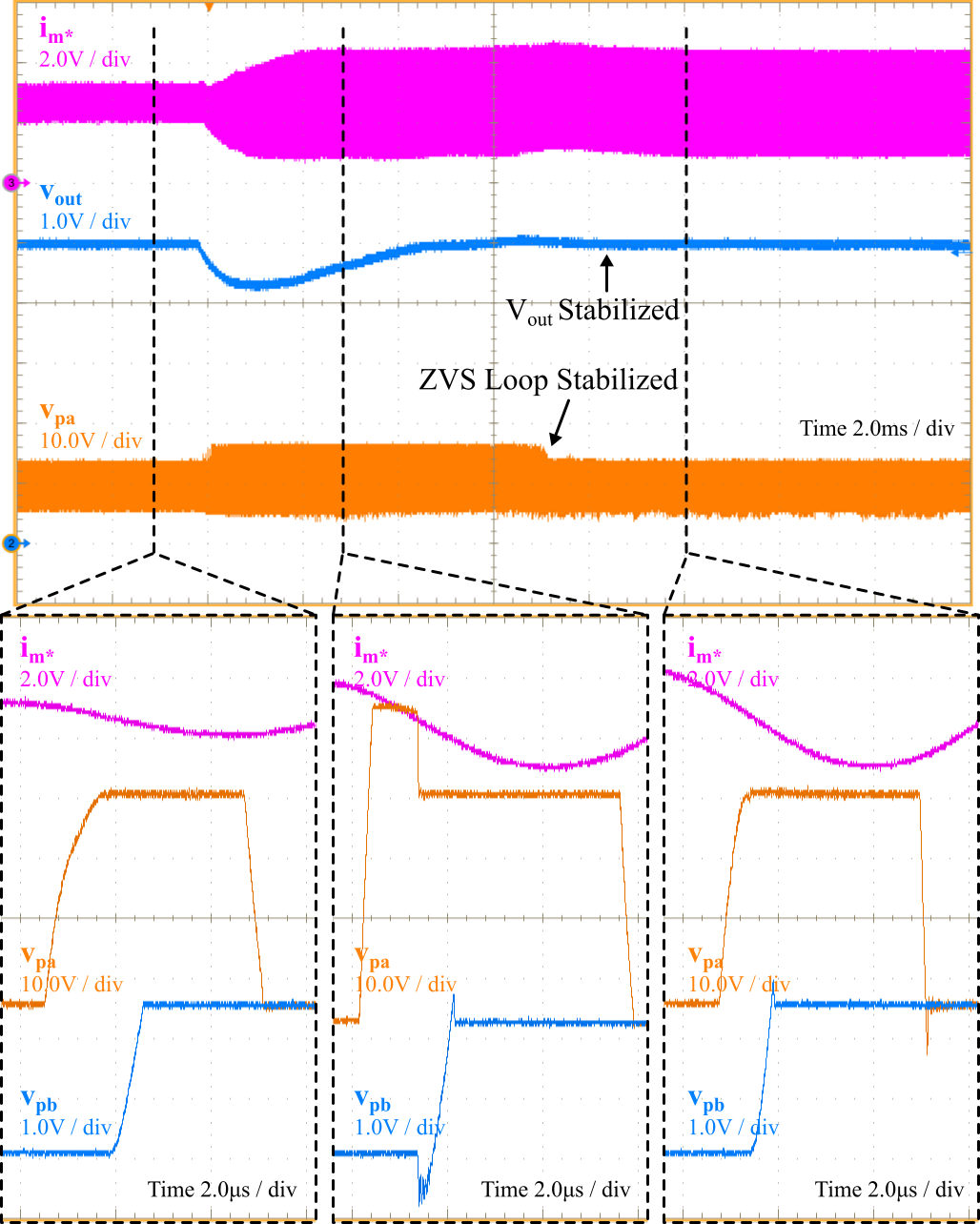}
    \caption{Transient response of the converter with PT-I as the piezoelectric component. $R_{load}$ decreases from $324\Omega$ to $75\Omega$. $V_{in} = 12{\rm V}$, $V_{out} = 5{\rm V}$.\label{fig:load-transient}}
\end{figure}

\subsection{Start-up}
The start-up waveform is shown in Fig.~\ref{fig:startup}. Both the converter and the FPGA controller board are powered before the start-up process but the state machine is initially paused (all switches are turned off). After the state machine starts at stage $S_{12}$, the capacitors in the phase-shifting circuit takes some time to reach equilibrium, during which the restored motional current signal is beneath the hysteresis of the comparator, thus the switching sequence is first driven by the pre-set maximum time span of each stage.

When the phase-shifting circuit reaches the steady state and the restored motional current gains enough amplitude, the zero-crossing events are triggered. The switching sequence is then synchronized in phase and allows the PI loop to drive $V_{out}$ to the reference steadily. Simultaneously, $T_{56}$ is gradually reduced by the ZVS loop until a certain point shown in lower Fig.~\ref{fig:startup}, when the overshoot of $v_{pa}$ at $t_{6B}$ is eliminated. This verifies the functionalities of the binary feedback loop. Eventually the system settles as shown in Fig.~\ref{fig:operation-waveform} and ZVS is achieved on all transitions. The complete process starting from the start-up of the state machine took 8.6 ms to converge.

\subsection{Load Transient}
The results of the load transient experiment is shown in Fig.~\ref{fig:load-transient}. As the load resistance stepped down from $324\Omega$ to $75\Omega$, and the voltage re-stabilized in 4.4 ms with an undershoot of 0.7V. Similar to the start-up, the experiment validates the voltage regulation loop as well as the ZVS loop.

\section{Conclusion}
This work demonstrates the feasibility of using ring-dot PT structure to measure the primary side motional current of piezoelectric resonators under radial vibration mode. Based on this, a simplified control strategy consists of a state machine, a PI loop, a binary accumulator, five comparators and a low-speed ADC is proposed and verified on a PR-based non-isolated converter prototype. Experiment results reveal that all transitions during a switching sequence realize ZVS, two of them are also ZCS. Thus the switching loss is minimized and the converter achieves 96.6\% peak efficiency. The significant reduction on sophisticated hardware and software reliance makes the large-scale application of the piezoelectric-based magnetic-less converters more practical.

Furthermore, the methodology of applying sensing electrodes to measure the motional current may be applied to multi-port PTs and other vibration modes at higher frequency, though the configuration of the electrodes should be studied case-by-case. The proposed control strategy is also capable of high frequency operation of several MHz. 

Nevertheless, there are still several issues that need to be addressed. The propagation delays of the feedback chain must be shortened to better adapt to higher frequency, of which system integration may be possible solution~\cite{koHybridPiezoelectricResonatorbased2026}. There should also be an advanced packaging technique that does not interfere the vibration while providing fixation and heat dissipation, especially because the extra sensing electrodes result in more contacts. Lastly, the reliability and aging properties of the piezoelectric-based converters is still be verified. They will be included in our future research.

{\appendices
\section{Derivation of the Steady-State Current-Sensing Transfer Function\label{appendix:derivation}}

\begin{figure*}
    \begin{subequations}
        \setcounter{equation}{1}
        \begin{equation}
            {\bf{M}} = \left( {\begin{array}{*{20}{c}}
            {{Z_{i2}} + {Z_{i3}} + {Z_{m1}} + {Z_{m3}}}&{ - {Z_{m3}}}&0&0&0&0&0\\
            { - {Z_{m3}}}&{{Z_{m2}} + {Z_{m3}} + {Z_{o1}}}&{{Z_{o2}}}&0&0&0&0\\
            0&{ - {Z_{o3}}}&{{Z_{o2}} + {Z_{o3}}}&{ - A}&0&0&0\\
            0&0&0&{j\omega \left( {{C_{02}} + {C_{lk}}} \right)}&0&0&{ - 1}\\
            0&0&0&{j\omega {C_{lk}}}&1&{ - 1}&0\\
            { - A}&0&0&0&0&1&0\\
            0&{ - A}&A&0&0&0&1
            \end{array}} \right)
        \end{equation} 
    \end{subequations}
    \addtocounter{equation}{-1}
    \hrule
\end{figure*}

The parameters of the electromechanical model in Fig.~\ref{fig:ring-dot-equivalent-circuit} are calculated by~\eqref{eq:mechanical-impedance},~\eqref{eq:static-capacitance}, and~\eqref{eq:current-to-velocity}~\cite{forresterCircuitSimulatorCompatible2023a}.

\begin{subequations}
    \begin{align}
        {Z_1} &= \frac{\alpha}{r_{in}}  \cdot \left[ {\begin{array}{*{20}{l}}
        {\pi {r_{out}}\beta (\upsilon {r_{in}}){J_1}(\upsilon {r_{out}})}\\
        { - \pi {r_{out}}\gamma (\upsilon {r_{in}}){Y_1}(\upsilon {r_{out}}) - 2{r_{in}}}
        \end{array}} \right]
        \\
        {Z_2} &= \frac{\alpha}{r_{out}}  \cdot \left[ {\begin{array}{*{20}{l}}
        {\pi {r_{in}}\beta (\upsilon {r_{out}}){J_1}(\upsilon {r_{in}})}\\
        { - \pi {r_{in}}\gamma (\upsilon {r_{out}}){Y_1}(\upsilon {r_{in}}) - 2{r_{out}}}
        \end{array}} \right]
        \\
        {Z_3} &= 2\alpha
        \\
        \alpha  &= \frac{{2jh/\left[ {s_{11}^x\omega \left( {{\sigma ^x}^2 - 1} \right){r_{in}}{r_{out}}} \right]}}{{{Y_1}(\upsilon {r_{out}}){J_1}(\upsilon {r_{in}}) - {Y_1}(\upsilon {r_{in}}){J_1}(\upsilon {r_{out}})}}
        \\
        \beta (x) &= \left[ {\left( {{\sigma ^x} - 1} \right){Y_1}(x) + x{Y_0}(x)} \right]
        \\
        \gamma (x) &= \left[ {\left( {{\sigma ^x} - 1} \right){J_1}(x) + x{J_0}(x)} \right]
        \\
        \upsilon  &= \omega \sqrt {\rho s_{11}^x\left( {1 - {\sigma ^x}^2} \right)}
        \\
        {\sigma ^x} &=  - \frac{{s_{12}^x}}{{s_{11}^x}}
    \end{align}
    \label{eq:mechanical-impedance}
\end{subequations}

\begin{subequations}
    \begin{align}
        {C_{0}} &= \frac{{\pi \left( {r_b^2 - r_a^2} \right)\left( {\varepsilon _{33}^T - 2d_{31}^2/\left[ {s_{11}^E\left( {1 - \sigma } \right)} \right]} \right)}}{h}\\
        {C_{02}} &= \frac{{r_c^2 - r_b^2}}{{r_b^2 - r_a^2}}{C_0}
    \end{align}
    \label{eq:static-capacitance}
\end{subequations}

\begin{equation}
    A = \frac{{2\pi {d_{31}}}}{{s_{11}^E\left( {1 - {\sigma^E} } \right)}}
    \label{eq:current-to-velocity}
\end{equation}

Among the physical constants, $s^{x}$ is the elastic compliance tensor. $s^{x} = s^{E}$ if the ring is covered with electrodes, and $s^{x} = s^{D}$ if not. $s^{x}$ becomes more complicated if only one side of the ring is covered, yet it is found that $s^{D}$ is still an acceptable approximation. Likewise, the Poisson's ratio of the covered regions $\sigma^E$ and that of the uncovered regions $\sigma^D$ are used correspondingly.

The complete electromechanical model in Fig.~\ref{fig:ring-dot-equivalent-circuit} allows applying circuit analysis methods to examine the aforementioned properties of the PT. The circuit equations are written as~\eqref{eq:circuit-matrix}.

\begin{subequations}
    \setcounter{equation}{0}
    \begin{equation}
        {\bf{M}}\left( {\begin{array}{*{20}{c}}
        {{v_a}{r_a}}\\
        {{v_b}{r_b}}\\
        {{v_c}{r_c}}\\
        {{V_{sen}}}\\
        {{I_{p}}}\\
        {{I_{m}}}\\
        {{I_{m2}}}
        \end{array}} \right) = \left( {\begin{array}{*{20}{c}}
        {A{V_{p}}}\\
        0\\
        0\\
        {{V_{p}}j\omega {C_{lk}} - {I_{sen}}}\\
        {{V_{p}}j\omega \left( {{C_{lk}} + {C_{0}}} \right)}\\
        0\\
        0
        \end{array}} \right)
    \end{equation}
    \label{eq:circuit-matrix}
\end{subequations}

Subsequently, the transfer function between the primary side motional current and the secondary side sensing voltage is derived as~\eqref{eq:vout-irin}.

\begin{subequations}
    \begin{align}
        &G = \frac{{{V_{sen}}}}{{{I_m}}} = \frac{{{A^2}{V_{p}}\Omega _1^2 + \left( {j\omega {C_{lk}}{V_{p}} - {I_{sen}}} \right)\Omega _2^3}}{{{A^2}\left[ {\begin{array}{*{20}{l}}
        {{A^2}{V_{p}}\Omega _3^1 + j\omega {V_{p}}\left( {{C_{lk}}\Omega _4^2 + {C_{02}}\Omega _5^2} \right)}\\
        { + {I_{sen}}\Omega _1^2}
        \end{array}} \right]}}
        \\
        &\Omega _1^2 = {Z_{m3}}{Z_{o2}}\\
        &\Omega _2^3 = \left( {{Z_{i2}} + {Z_{i3}} + {Z_{m1}}} \right)\Omega _5^2 + {Z_{m3}}\left( {\Omega _4^2 - {Z_{m3}}{Z_{o3}}} \right)\\
        &\Omega _3^1 = {Z_{m2}} + {Z_{m3}} + {Z_{o1}} + {Z_{o2}}\\
        &\Omega _4^2 = \begin{array}{*{20}{c}}
        {{Z_{m2}}{Z_{o2}} + {Z_{m2}}{Z_{o3}} + {Z_{m3}}{Z_{o3}}}\\
        { + {Z_{o1}}{Z_{o2}} + {Z_{o1}}{Z_{o3}} + {Z_{o2}}{Z_{o3}}}
        \end{array}\\
        &\Omega _5^2 = \Omega _1^2 + \Omega _4^2
    \end{align}
    \label{eq:vout-irin}
\end{subequations}
which is referred to as the {\it current-sensing transfer function} in this work. $\Omega^n$ denotes a n-th degree polynomial of mechanical impedance. 

To have an analytical insight of~\eqref{eq:vout-irin}, it is assumed here that the middle section of the PT is negligibly small so that $Z_{m[1,2]} \rightarrow 0$, $Z_{m3} \rightarrow \infty$. Besides, the load impedance of the secondary side is assumed to be infinitely large to make $I_{sen} \rightarrow 0$. Then, the current-sensing transfer function is simplified as~\eqref{eq:tf-simplified}.

\begin{subequations}
    \begin{align}
        &G' = \mathop {\lim }\limits_{{Z_{m[1,2]}}, I_{sen} \to 0} \mathop {\lim }\limits_{{Z_{m3}} \to \infty } \frac{{{V_{sen}}}}{{{I_m}}}
        \nonumber
        \\
        &= \frac{{{A^2}{Z_{o2}} + j\omega {C_{lk}}\Omega_6^2 }}{{{A^2}\left[ {{A^2} + j\omega {C_{lk}}{Z_{o3}} + j\omega {C_{02}}\left( {{Z_{o2}} + {Z_{o3}}} \right)} \right]}}
        \\
        &\Omega_6^2  = \left( {{Z_{i2}} + {Z_{i3}} + {Z_{o1}}} \right)\left( {{Z_{o2}} + {Z_{o3}}} \right) + {Z_{o2}}{Z_{o3}}
    \end{align}
    \label{eq:tf-simplified}
\end{subequations}

Assuming that $C_{lk} \rightarrow 0$, the current-sensing transfer function can be further simplified as~\eqref{eq:tf-further-simplified}.

\begin{equation}
    G'' = \mathop {\lim }\limits_{{C_{lk}} \to 0} G = \frac{{{Z_{o2}}}}{{{A^2} + j\omega {C_{02}}\left( {{Z_{o2}} + {Z_{o3}}} \right)}}
    \label{eq:tf-further-simplified}
\end{equation}

Firstly, the phase of the current-sensing transfer function near DC frequency is found by~\cite{olverNISTHandbookMathematical2010},

\begin{align}
    &\mathop {\lim }\limits_{\omega  \to 0} j\omega {C_{02}}\left( {{Z_{o2}} + {Z_{o3}}} \right)
    \nonumber
    \\
    &= \frac{{2\pi h{C_{02}}}}{{s_{11}^E\left( {1 - {\sigma ^E}^2} \right)}} \cdot \frac{{\left( {{\sigma ^E} + 1} \right)r_c^2 - \left( {{\sigma ^E} - 1} \right)r_b^2}}{{\left( {r_c^2 - r_b^2} \right)r_c^2}}
    \label{eq:jwC02-limit}
\end{align}

For PZT materials, ${\sigma^E} \approx 0.26 \sim 0.34$~\cite{erhartPiezoelectricCeramicResonators2017}, thus~\eqref{eq:jwC02-limit} is always positive. Besides, $Im(Z_{o2}) > 0, Re(Z_{o2}) = 0$ by defination. Therefore, the phase of~\eqref{eq:tf-further-simplified} at low frequency is always $90^\circ$:

\begin{equation}
    \mathop {\lim }\limits_{\omega  \to 0} \phi (G'') = \frac{\pi }{2}
    \label{eq:low-freq-phase}
\end{equation}

Secondly, \eqref{eq:tf-simplified} reveals that the current-sensing transfer function has an infinite number of poles because of the Bessel functions and the phase of it can either be $90^\circ$ or $-90^\circ$ if mechanical loss is ignored. It is of interest to study when the phase will flip to $-90^\circ$ and whether one of the flipping points is near the common operating frequencies.~\eqref{eq:tf-further-simplified} shows that the first pole must be at a frequency higher than $\omega_0$ which makes $Z_{o2} + Z_{o3} = 0$. Let $\upsilon_0$ be the corresponding wave number of $\omega_0$ defined in~\eqref{eq:mechanical-impedance} and extend $Z_{o2}$, $Z_{o3}$. $\upsilon_0$ should be the first solution of:
\begin{equation}
    \frac{{\beta \left( {\upsilon {r_c}} \right)}}{{\gamma \left( {\upsilon {r_c}} \right)}} = \frac{{{Y_1}({\upsilon}{r_b})}}{{{J_1}({\upsilon}{r_b})}}
\end{equation}
for $\upsilon_0 r_c >> 1$, the Bessel functions collapse to trigonometric functions~\cite{olverNISTHandbookMathematical2010}:
\begin{equation}
    \upsilon \left( {{r_c} - {r_b}} \right) = \arctan \frac{{{\sigma ^E} - 1}}{{{\upsilon _0}{r_c}}} + \left( {\frac{1}{2} + n} \right)\pi 
\end{equation}
the first solution can be found with the Taylor series with $n=0$:
\begin{equation}
    {\upsilon _0} \approx \frac{{\frac{\pi }{2} + \sqrt {\frac{{{\pi ^2}}}{4} - \frac{{4\left( {{r_c} - {r_b}} \right)}}{{{r_c}}}\left( {1 - {\sigma^E} } \right)} }}{{2\left( {{r_c} - {r_b}} \right)}} \approx \frac{\pi }{{2\left( {{r_c} - {r_b}} \right)}}
    \label{eq:flipping-point}
\end{equation}

On the other hand, the ideal operation frequency is near the first resonant frequency $\omega_{r1}$~\cite{bolesEnumerationAnalysisDC2021a}. Let $\upsilon_{r1}$ be the corresponding wave number, it can be approximated by the first solution of~\eqref{eq:resonant-point}~\cite{erhartPiezoelectricCeramicResonators2017, forresterCircuitSimulatorCompatible2023a}.

\begin{equation}
    \gamma \left( \upsilon_{r1} {r_c} \right) = 0
    \label{eq:resonant-point}
\end{equation}
when simplified under the assumption of $\upsilon_{r1}r_c >> 1$:
\begin{equation}
    {\upsilon _{r1}} \approx \frac{{\frac{{3\pi }}{4} + \sqrt {{{\left( {\frac{{3\pi }}{4}} \right)}^2} - 4\left( {1 - {\sigma ^E}} \right)} }}{{2{r_c}}} \approx \frac{{3\pi }}{{4{r_c}}} < \upsilon_0
    \label{eq:resonant-frequency}
\end{equation}

Therefore, the flipping point of the current-sensing transfer function is much higher than the first resonant frequency of the PR. 

By combining~\eqref{eq:vout-irin},~\eqref{eq:low-freq-phase} and~\eqref{eq:resonant-frequency}, it is found that the steady state current-sensing transfer function under ideal condition satisfies:
\begin{subequations}
    \begin{align}
        \phi \left( {\frac{{{V_{sen}}}}{{{I_m}}}} \right) &\equiv \frac{\pi }{2} \quad \left( {\upsilon  < {\upsilon _{0}}} \right)
        \\
        \left| {{V_{sen}}} \right| &\propto \left| {{I_m}} \right|
    \end{align}
\end{subequations}
which fulfills the requirements of current-sensing.

Notably, the electromechanical model in Fig.~\ref{fig:ring-dot-equivalent-circuit} and~\eqref{eq:mechanical-impedance} is derived under the assumption of harmonic excitation containing $\omega \in (0, \infty)$. As a result, all conclusions obtained from this model are only valid under steady state.
}

\bibliographystyle{IEEEtran}
\bibliography{Reference}

\end{document}